\documentclass[a4paper,11pt]{article}
\usepackage{jcappub}
\input{epsf}
\newcommand{\be}{\begin{eqnarray}}
\newcommand{\ee}{\end{eqnarray}}
\notoc
\begin{document}

\title{Beyond the standard $\Lambda$CDM cosmology: the observed
structure of DM halos and the shape of the power spectrum}
\author[a, b]{M. Demia\'nski}
\author[c, d]{A.G. Doroshkevich}
\affiliation[ a]{
Institute of Theoretical Physics, University of Warsaw,
02-093 Warsaw, Poland}
\affiliation[b ]{ Department of Astronomy, Williams College,
Williamstown, MA 01267, USA}
\affiliation[ c]{ Lebedev Physical Institute of the
Russian Academy of Sciences,117997, Moscow,  Russia}
\affiliation[ d ]{ National Research Center Kurchatov Institute,
123182, Moscow,  Russia}
\emailAdd{Marek.Demianski@fuw.edu.edu}
\emailAdd{dorr@asc.rssi.ru}
\date{Received 2016  .../ Accepted ...,     }

\abstract{
Recent advances in observational astronomy allow to
study various groups of Dark Matter (DM) dominated objects from
the dwarf spheroidal (dSph) galaxies to clusters of galaxies
that span the mass range from $10^{6} M_{\odot}$ to $10^{15} M_{\odot}$.
To analyze data of this divers collection of objects we used
a simple toy model of spherical DM halo formation that was
initially proposed by Peebles. This model introduces the concept
of the epoch or redshift of halo formation. Using this concept
we analyzed selected sample of DM dominated objects and we have found
empirical correlations between the virial mass, $M_{vir}$,
of halos and basic parameters of their cores, namely, the
mean DM density, pressure and entropy. These correlations are
a natural result of similar evolution of all such objects.
It is driven mainly by gravitational interactions of DM what 
implies a high degree of self similarity of both the process
of halos formation and their internal structure.

We confirm the CDM--like shape of both the small and large scale 
power spectrum. However, our reconstruction of the
evolutionary history of observed objects differs from expectations 
of the standard $\Lambda$CDM cosmology and requires
either a multicomponent composition of DM or a more complex 
primordial power spectrum of density perturbations with
significant excess of power at scales of clusters of galaxies 
and larger. We demonstrate that a model with suitable
combination of the heavy DM particles (CDM) and DM particles 
with large damping scale (HDM) could provide a successful
description of the observational data in a wide range of masses.}

\keywords{cosmology: formation of DM halos, galaxies and
clusters of galaxies -- initial power spectrum -- composition
of dark matter.
}

\maketitle

\section{Introduction}

Observations of the CMB fluctuations by the WMAP mission and
ground based telescopes established the $\Lambda$CDM model
as the best cosmological model (Bennet et al. 2003; Komatsu
2011; Larson 2011; Saro 2014). This inference was supported
by the Planck measurements (Ade et al. 2016), observations
of clusters of galaxies (see, e.g., Burenin \& Vikhlinin
2012) and baryonic acoustic oscillations (Eisenstein \& Hu
1998; Meiksin et al. 1999; Samushia et al., 2014). These
observations are consistent with the present day theoretical
expectations (Ellis et al. 2016; see, however Rubakov
2014),  but other important problems remain unsolved. Thus,
we do not know the composition and basic properties of dark
matter (DM) and dark energy, and we have no information
about the possible ``missing'' cosmological parameters that
allow to combine the complex small scale spatial matter
distribution with the surprisingly high apparent spherical
symmetry and large scale homogeneity of the observed Universe.

Indeed, attempts of direct detection of DM particles by
DAMA (Bernabei 2010), CRESST-II (Angloher et al. 2013),
and SuperCDMS (Agnese 2014) experiments and other (see
reviews in Mayet et al., 2016; Blennow 2016) have
not yet produced reliable positive results. Hence up to now
we have no reliable estimates of the mass, the nature, and
properties of DM particles (see, e.g., Blinnikov 2014;
Abdallah et al., 2015; Buchmueller et al. 2015; Borsanyi et
al., 2016).

Now the main hope of identifying the DM composition
is based on the recent progress
in observations of objects at high redshifts (Robertson
et al. 2015; McLeod 2015; Bouwens et al. 2015a,b; Mitra
et al. 2015; Zitrin, 2015) and on observations of Large
Scale Structure elements with a size up to 1700 Mpc
(see, e.g., Einasto 2014; Balazs et al. 2015; Horvath et al.
2015). These observations are connected with the unexpected
properties of the initial perturbations and, so, both with
the inflation models and interpretation of the CMB
observations (see, e.g. discussions in Demia\'nski \&
Doroshkevich 2007; Doroshkevich \& Verkhodanov 2011).

Results of observations of the small scale structure of the
Universe are not so crucial and they require only moderate
corrections of the present day models (see review of
Berezinsky et al., 2014). Thus, the emerging conflict between
the Standard $\Lambda$CDM model and observations of clustering
on subgalactic scales is widely discussed. It is
believed that the standard $\Lambda$CDM model predicts an
excess of low--mass satellites of galaxies similar to the
Milky Way (missing satellite problem, see, e.g., Moore et al.
1999; Klypin et al. 1999; Bovil \& Ricotti 2009; Trujillo-Gomez
et al. 2011). Simulations show that this problem does not
appear in the WDM models, which however cannot reproduce
other observations and so cannot be considered as the
basic cosmological model. Brook,\,\&\,Cintio (2015)
propose more complex density profiles while last publications
(Chan et al. 2015; Sawala et al. 2016) maintain that the high
resolution simulations correctly describe the evolution of
baryonic component and can eliminate disagreements with
observations. In this paper we show (Sec.7) that this problem
is also alleviated in the models with more complex power
spectrum of density perturbations.

 During last years the formation and evolution of the first dwarf
galaxies is discussed very actively. In particular, Weisz
et al. (2014) considered the star formation history for 40
observed galaxies of the Local Volume, Kirby et al. (2017) discuss
in details the possible evolution of three galaxies -- Leo A,
Aquarius \& Sagittarius DIG --- and conclude that they are DM
dominated, Fitts et al. (2016) presented set of simulations
of star formation and demonstrated the great diversity of
the possible histories. This list can be extended.

It is clear that if the baryonic component -- gas and stars -- determine
properties of many observed galaxies then it could be also
important for low mass galaxies considered now as the DM
dominated ones. It is specially alluring to use baryons to explain
the core--cusp problem which is seen as a discrepancy between
the observed and simulated shape of the density profiles in central
regions of relaxed objects (see, e.g., Bovill \& Ricotti, 2009;
Koposov et al., 2009; Walker \& Penarrubia, 2011; Boylan--Kolchin
et al., 2012; Penarrubia et al 2012; Governato et al. 2012; Sawala,
2013; Teyssier et al. 2013; Laporte et al. 2013; Collins et al.
2014; Brook~\&~Cintio 2015; Sawala et al. 2016). In particular
the strong possible concentration of stars in the centers of galaxies
is often discussed as a possible way of transformation of DM cusp
into the core. Of course some time this is possible but it is
not obligatory.

The interest to this problem is concentrated at the evolution of
core itself as any of its impact on the properties of galaxies as
a whole is limited. In some aspects this problem is similar to the
problems of AGNs or the super massive black holes that are
some time also situated in the centres of galaxies. Indeed both the DM virial mass
and outer structure of galaxies or DM halos are weakly sensitive to
the composition, shape or other properties of the core. Thus the
Navarro -- Frenk -- White (Navarro et al. 1995, 1996, 1997) and
Burkert (1995) density profiles are very different near the
center but the difference becomes moderate outside the core.

These results indicate that many factors influence the evolution
of baryons and specially the star formation rate. But the significance
of discrepancies between the observed and simulated properties is
strongly overestimated as in these cases properties determined by
methods of very different reliabilities are compared.

In fact properties of simulated objects are calculated using
averaged characteristics (3D density, velocity dispersion etc.) of
many particles of DM component. In contrast, the same properties
of observed galaxies are derived from measurements of 2D velocity
dispersions of only limited number of stars. In turn the process of
formation of stars is sensitive to heating and cooling of the
gaseous component, to creation and dissipation of macroscopic
turbulent motions and so on. In particular these factors are
responsible for the formation of discs, bulges, clusters of stars
and other elements of the complex internal structure of the observed
massive galaxies. The phase density and entropy of DM, stars
and other components of galaxies are different and so the final
density profiles depend upon the composition of the objects. However
now this complex problem is far from conclusive solution.

In this paper we are interested in discussion of observed properties
of DM dominated objects and thus to reduce the possible impact
of baryonic component we use in our analysis objects with limited
contribution of baryons (dSph and ultra diffused galaxies
(UDG), small fraction of groups and clusters of galaxies, and some
of the LSB galaxies). Next we analyze the mean properties
of the cores rather than the more problematic properties of the
centers or of the virial masses. Finally we reject all objects with
calculated properties which strongly deviate from our expectations.
This rejection decreases the final sample but it makes it more
homogeneous. More detailed discussion is presented in Sec. 2. 

Limited reliability of both discussed contradictions is
enhanced by the limited number of observations
and simulations (see, e.g., Mikheeva, Doroshkevich, Lukash
2007; Doroshkevich, Lukash \& Mikheeva 2012; Pilipenko et
al. 2012; Newman et al. 2013; Brook et al., 2014; Miller
et al. 2014;  Brooks \& Zolotov 2014; Breddels \& Helmi
2014; Nipoti \& Binney, 2015; Mamon et al. 2015). Among
the discussed discrepancies the ``abundance matching
relationship'' (Garrison-Kimmel et al. 2014c; Brook et al.
2014) and the 'Too Big to Fail' (TBTF) problem (Boylan-Kolchin
et al. 2012; Tollerud et al. 2014; Garrison-Kimmel 2014a,b;
Klypin et al. 2015) are the most often discussed and most important.
The TBTF problem is discussed in section 7.2 in more details.

Other possible corrections of the standard $\Lambda$CDM
cosmological model are discussed in the context of small
divergences of the local and CMB measurements of the
cosmological parameters. Thus, emerging tensions between
different measurements of the Hubble constant can be explained
by introduction of an unstable fraction of DM particles
(Berezhiani et al. 2016). The model with unstable DM
particles is also discussed by Wang (2014) and Enqvist
et al. (2015). These hypotheses reanimate similar earlier
proposed models (Turner et al. 1984; Doroshkevich \&
Khlopov 1984; Doroshkevich et al. 1986, 1988). New
observational constrains of possible decay of DM
particles are discussed by Baring et al. (2016).
Other original models of DM component were discussed
by Anderhalsen et al. 2013 and Schewtchenko et al.
2015.

Now the sterile neutrinos with a various masses are widely
discussed as the popular candidate for the DM particles (see,
e.g., Gorbunov 2014). More popular are the sterile neutrinos
in the keV range (see, e.g., reviews of Feng 2010;
Boyarsky et al., 2009c, 2013; Kusenko 2009; Marcovi$\breve{c}$
\& Viel 2013; Pontzen \& Governato 2014; Horiuchi et
al., 2015). On the other hand models of light sterile
neutrinos are also quite popular (see, e.g., Abazajian 2012;
Kopp et al., 2013; Zysina, Fomichev \& Khruschov 2014). In
some of these complex models sterile neutrinos do not reach
the state of thermal equilibrium. A possible decay of sterile
neutrinos is also discussed (e.g., Ferrer \& Hunter 2013;
Bulbul et al. 2014; Enqvist et al. 2015). These models imply
a multicomponent composition of DM (Battye \& Moss 2014;
Whittaker et al. 2014, 2015; Battye et al. 2015; Mirizzi et
al. 2015; Abbott 2015).  However, recently properties of
sterile neutrinos were strongly constrained by the Ice Cube
observations (Aartsen et al. 2016).

Observations of DM dominated halos are very well complemented
by numerical simulations that allow to trace and
investigate the early stages of halos evolution, as well as
the process of halos virialization and formation of their
internal structure. The formation of virialized DM halos begins
as the anisotropic collapse in accordance with the
Zel'dovich theory of gravitational instability (Zel'dovich
1970, Zel'dovich \& Novikov, 1983; Demia\'nski et al. 2011).
During later stages the evolution of such objects becomes more
complex because it is influenced by their anisotropic
environment and anisotropic matter inflow (see, e.g., real
cluster representations in Pratt et al. 2009). Moreover all
the time the evolution of halos is unavoidably accompanied
by the process of violent relaxation and merging, which is
also well reproduced by simulations.

Analysis of simulated halos can be performed in a wide range
of halo masses and redshifts, what allows to improve the
description of properties of relaxed halos of galactic and
cluster scales and to link them with the power spectrum of
initial perturbations. Thus, it is established that after a
period of rapid evolution the main characteristics of
majority of the high density DM cores of virialized halos
become frozen and their properties are only slightly changing
owing to the accretion of diffuse matter and/or the evolution
of their baryonic component. The basic properties of the
relaxed DM halos in the framework of the $\Lambda$CDM model
were investigated in many papers (see, e.g., Tasitsiomi et
al. 2004; Nagai et al. 2007; Croston et al. 2008; Pratt et al.,
2009, 2010; Vikhlinin et al. 2006, 2009; Arnaud et al.
2010; Klypin et al. 2011; Kravtsov \& Borgani 2012; Ruffini
et al. 2015).

Properties of halos formed by collisionless DM particles are
determined by the process of violent relaxation only, what
leads to the self similarity of the internal structure of halos
and implies that halo properties, cosmological
evolution and initial perturbations are correlated.  Thus, for
the WDM model the available simulations (see, e.g.,
Maccio 2012, 2013; Angulo, Hahn, Abel, 2013; Schneider, Smith \&
Reed, 2013; Wang et al. 2013; Libeskind et al. 2013;
Marcovi$\breve{c}$ \& Viel 2013; Schultz et al. 2014; Schneider
et al. 2014; Dutton et al. 2014) show that in
accordance with expectations the number of low mass halos
decreases and the central cusp in the density profile is
transformed into the core. For larger halos the standard density
profile is formed again, but formation of high density
objects is accompanied by appearance of some unexpected phenomena.
Thus, Maccio et al. 2012 \& 2013, Schneider et al.
2014 confirm the decrease of matter concentration in halos in
the WDM model in comparison with the CDM model, but they
conclude that the 'standard' WDM model is not able to reproduce
the density profiles of low mass galaxies. This
inference is enhanced by Libeskind et al. 2013, where in
contrast with the CDM model their simulation with 1 keV WDM
particles cannot reproduce the formation of the Local Group. In
turn, Schultz et al. 2014 note that in their
simulations with ~3keV WDM particles formation of objects at
large redshifts and reionization are over suppressed. This
means that the simulations of WDM and possibly the multicomponent
DM models require further detailed analysis and it is
necessary to put special attention to reproduce links between
the mass function of halos and the power spectra with the
free streaming cut-off.

Without doubt, one can expect a rapid progress in simulations
of more complex models of the process of structure
formation. However, for preliminary discussion of such models we
used the semi analytical description of DM dominated
objects proposed in our previous paper (Demia\'nski \&
Doroshkevich 2014). It is based on the approximate analytical
description of the structure of collapsed halos formed by
collisionless DM particles. During the last fifty years
similar toy models have been considered many times in order to
study various aspects of the nonlinear process of
condensation of matter (see, e.g., Peebles 1967, 1980; Zel'dovich
\& Novikov 1983; Fillmore and Goldreich 1984;
Gurevich \& Zybin 1995; Bryan \& Norman 1998; Lithwick \& Dalal
2011). Of course such models ignore many important
features of the process of halos formation and are based on
the assumption that a virialized DM halo is formed during a
short period of the spherical collapse at $z\approx z_{f}$ and
later on its parameters vary slowly owing to the
successive accretion of matter (see, e.g., discussion in
Bullock et al. 2001; Diemer et al. 2013; Diemer\,\&\,Kravtsov 2014).

In this paper we consider observations of DM dominated relaxed
objects -- dSph and UDG galaxies and selected sample of
19 clusters of galaxies  -- in a wide range of their masses,
$10^6\leq M_{vir}/M_\odot\leq 10^{14}$. These data
demonstrate the well known correlation between the mass of
virialized DM halo $M_{vir}$, the epoch (or redshift) of its
formation, $t_f,\,\&\,z_f$, and the corresponding mean DM density,
$\langle\rho_{vir}\rangle$. The epoch of galaxies
formation was roughly estimated already by Partridge \& Peebles
1967a,b. Later on more accurate estimates of this
epoch have been discussed in the framework of simple toy models
(Zeldovich\,\&\,Novikov 1983; Lacey \& Cole 1993; Bryan
\& Norman 1998).

Simulations also show that properties of the central cores of
virialized DM halos are mainly established during the
early period of halos formation and later on the slow pseudo --
evolution of cores dominates. This means that they are
less sensitive to actions of random factors that distort the
outer characteristics of observed objects (see, e.g.,
Diemer\,\&\,Kravtsov 2014). This means that analysis of the
central properties of DM dominated objects allows to obtain
more stable description of virialized objects. However, these
properties can be distorted by influence of the baryonic
component, what makes such analysis more complex.

Our results confirm the self similar character of evolution of
the DM dominated objects. Application of the excursion
set approach (Press, Schechter 1974; Bardeen et al. 1986; Bond
et al. 1991) demonstrates unexpected mass dependence of
the amplitude of power spectrum that can be caused by a complex
composition of DM particles and/or by a more complex
inflationary period.

However, potential of this approach should not be overestimated.
As usual we can only determine the probability of
formation of objects and therefore it has only statistical
character rather than strict constrains or prediction.
Moreover in our discussion we use observational data of only
limited quality and representativity. Thus, we can only use
a small number  of recently observed DM dominated objects,
unfortunately their observed characteristics -- even so
important as the virial mass -- are known only with very limited
precision and significant scatter. Nonetheless the
potential of the proposed approach is significant as it considers
objects in a very wide range of masses. We hope that
further accumulation of observational data and their comparison
with the high resolution simulations will essentially
improve presented results.

This paper is organized as follows: In section 2 a simple model
of formation of  DM halos is presented. In section 3
generation of entropy of DM particles is illustrated and in
section 4 an approximate analytical description of
properties of DM halos is presented. In section 5 this technique
is applied to eight observed samples of DM dominated
objects. The mean characteristics of these objects are summarized
in Table 1 and are discussed in section 6\,. In
section 7 the correlation between halo masses and the redshift
of formation is compared with expectations of the
standard $\Lambda$CDM cosmology and the actual shape of the
power spectrum is estimated. Discussion and conclusions can
be found in section 8\,.

\subsection{Cosmological parameters}

In this paper we  consider the spatially flat $\Lambda$
dominated model of the Universe with the Hubble parameter,
$H(z)$, the mean critical density $\langle\rho_{cr}\rangle$,
the mean density of non relativistic matter (dark matter
and baryons), $\langle\rho_m(z)\rangle$, and the mean density
and mean number density of baryons, $\langle\rho_b(z)
\rangle\,\&\,\langle n_b(z)\rangle$, given by Komatsu et al.
2011, Hinshaw et al. 2013:
\[
H^{2}(z) = H_0^2[\Omega_m(1+z)^3+\Omega_\Lambda],\quad H_0=
100h\,{\rm km/s/Mpc}\,,
\]
\be
\langle\rho_b(z)\rangle = {3H_0^2\over 8\pi G}\Omega_b(1+z)^3
\approx 4\cdot 10^{-31}(1+z)^3\Theta_b\frac{g}{cm^3} \,,
\label{basic}
\ee
\[
\langle\rho_m(z)\rangle =
2.5\cdot 10^{-30}(1+z)^3\Theta_m\frac{g}{cm^3}=
34(1+z)^3\Theta_m\frac{M_\odot}{kpc^3} \,,
\]
\[
\langle\rho_{cr}\rangle=\frac{3H^2}{8\pi G},\quad \Theta_m=
\frac{\Omega_mh^2}{0.12},\quad \Theta_b=\frac{\Omega_b
h^2}{0.02}\,.
\]
Here $\Omega_m=0.24\,\&\,\Omega_\Lambda=0.76$ are the mean
dimensionless density of non relativistic matter and dark
energy, $\Omega_b\approx 0.04$ and $h=0.7$ are the
dimensionless mean density of baryons, and the dimensionless
Hubble constant measured at the present epoch. Cosmological
parameters presented in the recent paper of the Planck
collaboration (Ade et al. 2016) slightly differ from
those used above.

For this model the evolution of perturbations can be
described with sufficient precision by the expression
\be
\delta\rho/\rho\propto B(z),\quad B^{-3}(z)\approx
\frac{1- \Omega_m+2.2\Omega_m(1+z)^3}{1+1.2\Omega_m}\,,
\label{Bz}
\ee
(Demia\'nski \& Doroshkevich, 1999, 2004, 2014; Demia\'nski
et al. 2011) and for $\Omega_m\approx 0.25$ we get
\be
B^{-1}(z)\approx\frac{1+z}{1.35}[1+1.44/(1+z)^3]^{1/3}\,.
\label{bbz}
\ee
For $z=0$ we have $B=1$ and for $z\geq 1,\,
B(z)$ is reproducing the exact function with accuracy
better than 90\%.
For $z\geq 1$ these relations simplify. Thus, for
the Hubble constant and the function $B(z)$ we get
\be
H^{-1}(z)\approx \frac{0.85\cdot 10^{18}s}{\sqrt{\Theta_m}
(1+z)^{3/2}},\quad B(z)\approx \frac{1.35}{1+z}\,.
\label{bzz}
\ee

\section{Physical model of halos formation}

It is commonly accepted that in the course of complex
nonlinear condensation the DM forms stable virialized
halos with a more or less standard density profile
and their typical mass is slowly increasing with time.
Numerical simulations show that the virialized DM halos
with various masses are formed from initial perturbations
after a short period of rapid complex evolution. Such
virialized objects are observed as clusters of galaxies,
isolated galaxies and/or as high density galaxies within
less dense clusters of galaxies, filaments, superclusters
or other elements of the Large Scale Structure of the
Universe.  As usual we will characterize such
virialized objects by their virial mass, $M_{vir}$, which
is the most popular basic characteristic of
objects in spite of the low precision of its observational
determination.

Evidently the halo formation is a deterministic process
and properties of virialized objects correlate with
properties of the initial perturbations. However, the complex
character of the process of halo formation destroys many
correlations and allows to reveal only some of them. Thus,
the impact of the baryonic component (and stars) is manifested
as its heating by shock waves, subsequent cooling and infall
into central regions. However, such processes are effective
only for objects with virial masses $10^8\leq M_{vir}/M_\odot\leq
10^{13}$, while for less massive dSph objects and very massive
clusters of galaxies they are not very important.
These differences are caused by variations of efficiency
of hydrodynamical and thermal processes such as the heating
and cooling of gas, the thermal instability and the process
of star formation.
These processes are slow for dwarf galaxies and clusters of
galaxies, their evolution is mainly determined by gravitational
interactions. For such objects the most stable characteristics are
the density profile of DM halos and the mean density, pressure
and entropy of their cores. It can be expected that these
characteristics are moderately distorted in the course of
halos formation and evolution and they can be linked with
properties of initial perturbations.

Properties of both simulated and observed DM dominated virialized
objects -- galaxies and clusters of galaxies -- are
usually described in the framework of spherical models such as
the Navarro -- Frenk -- White (NFW) proposal (Navarro et
al. 1995, 1996, 1997; Ludlow et al. 2013), isothermal or Burkert
1995 \& 2015 models. This approach allows to discuss
and to link together both the general parameters of a halo, such
as its virial mass, period of its formation, relations
between the thermal and gravitational energy, and its internal
properties such as the density profile, mean density and
entropy of its core etc.

In this paper, using this model, we consider the observed
properties of DM dominated virialized objects -- the dSph
galaxies and some set of clusters of galaxies -- in a wide
interval of masses $10^6\leq M_{vir}/M_\odot\leq 10^{15}$ and
we assume that all relaxed DM halos are described by the NFW
density profile. Our approach is based on three arguments:
 \begin{enumerate}
\item{} all objects are formed from initial perturbations
described by the same power spectrum,
\item{} the internal structure of all halos is determined by
the same processes of violent relaxation,
\item{} many numerical simulations support these inferences.
\end{enumerate}

This approach implies a high degree of self similarity
in mass dependence of the main characteristics of DM halos.
Of course, it ignores many details of the complex process
of halos formation and, in particular, it does not prevent
formation of low mass objects at any redshift. But it
allows to obtain a very simple, though rough, general
description of the process of DM halos formation and
introduces some hierarchy of formed objects.

This analysis also  reveals a possible deviations of the
main observed characteristics from expectations based on
standard assumptions about the DM composition and/or initial
power spectrum used in theoretical models and simulations.
An example of such deviation is the 'To Big to Fail' effect
(Boylan--Kolchin et al. 2012; Garrison--Kimmel et al.,
2014a,b; Tollerud et al. 2014; Klypin et al. 2015; Hellwing
et al. 2015; Brook, \& Cintio 2015), what indicates that the
usually accepted models should be improved.

\subsection{DM density in virialized halos}

In this paper we assume that all relaxed DM halos are described by the NFW
density profile
\be
\rho(x)=\frac{\rho_0}{x(1+x)^2},\quad x=r/r_s\,,
\label{nfw-d}
\ee
where $\rho_0(M_{vir}),\,\&\,r_s(M_{vir})$ are model
parameters. Using this density we get that
\[
M(r)=M_sf_m(r/r_s),\quad M_s=4\pi\rho_0r_s^3,\quad
M_{vir}=M_sf_m(c)\,,
\]
\be
 f_m(x)=ln(1+x)-x/(1+x),\quad c=R_{vir}/r_s\geq 3\,,
\label{nfw-m}
\ee
\[
M(r_s)=M_sf_m(1)\approx 0.2M_s\,,
\]
where $c$ is the concentration and $R_{vir}$ is the halo
virial radius.  For the mean density of halos,
$\langle\rho_{vir}\rangle$ and the mean density of their central
core $\langle\rho_s \rangle$ we get
\[
\langle\rho_{vir}\rangle=3M(R_{vir})/4\pi R_{vir}^3=
3\rho_0f_m(c)/c^3\,,
\]
\[
\langle\rho_s\rangle=3M(r_s)f_m(1)/4\pi r_s^3\approx 0.6\rho_0\,,
\]
and finally we have
\be
\langle\rho_s\rangle=5.4\langle\rho_{vir}\rangle\left(\frac{c}{3}
\right)^3\frac{1}{f_m(c)}\,.
\label{nfw-ms}
\ee

These relations link together the fundamental characteristics
of halos, namely, their mean density, concentration and masses. Thus,
the function $f_m(x)$ ($f_m(4)=0.8$ and $f_m(8)=1.3$) shows
that for the most interesting population of objects we can neglect
the differences between $M_{vir}$ and $M_s$ and accept that
\[
M_{vir}\approx M_s(1\pm 0.25)\approx 5M(r_s)(1\pm 0.25),
\quad 4\leq c\leq 8\,.
\]
These relations allow to roughly estimate parameters of the observed
objects in spite of scarcity of observational data.

\subsection{The redshift of halo formation}

For each mass of DM halo its formation is a complex process
extended in time, what causes some ambiguity in the halo
parameters such as its virial mass and the epoch or redshift
of formation (see, e.g. discussion in Partridge \& Peebles
1967a,b; Peebles 1980; and more recently in Diemand, Kuhlen
\& Madau 2007; Kravtsov \& Borgani 2012). The main stages
of this process can be investigated in details with numerical
simulations (see, e.g. Demia\'nski et al. 2011). The analytic
description of this process is however problematic.

For a virialized DM halo of mass $M_{vir}$ the redshift (or
epoch) of formation, $z_f$, and the corresponding mean DM
density $\langle\rho_{vir}\rangle$ were roughly estimated by
Partridge \& Peebles 1967a,b and were later determined
more accurately in the framework of the simple {\it
phenomenological} toy models (Zeldovich\,\&\,Novikov 1983;
Lacey \& Cole 1993; Bryan \& Norman 1998). According to
these models the virial density is proportional to the
mean density (\ref{basic}) at the moment of object formation,
\be \langle\rho_{vir}\rangle=18\pi^2\langle\rho_m(z_f)
\rangle\approx\rho_{200}(1+z_f)^3\,, \label{rmod}
\ee
\[
\rho_{200}=200\langle\rho_m(0)\rangle=
0.68\cdot 10^4M_\odot/kpc^3=5\cdot 10^{-28}g/cm^3.
\]
In the Lacey -- Cole model halos are considered as formed
at the moment of collapse of homogeneous spherical DM
clouds and they are described by the adiabatic Emden model
with $\gamma=5/3,\,n=3/2$ (Peebles 1980; Zel'dovich\,\&\,
Novikov 1983).

The advantage of this approach is its apparent universality
and reasonable results that are obtained for massive clusters
of galaxies. However, both its precision and range of
applicability are limited owing to noted simplified
assumptions. Thus, even the numerical coefficient $18\pi^2$
is determined by the Emden model for halo description.
An important, but usually ignored, special feature of the Lacey
- Cole model is the strong mass dependence of the virial density,
\be
\langle\rho_{vir}\rangle\propto M_{vir}^{-2} \,.
\label{mvir}
\ee

Attempts to use these relations for unified description of
observed DM halos of galactic and clusters scales
immediately leads to strongly unacceptable results. Thus,
expectations of the model (\ref{rmod},\,\ref{mvir}) are in
contrast with the weak mass dependence of the observed
parameters of DM dominated objects (sections 5\,\&\,6). It is
also important that for dSph galaxies Eq. (\ref{rmod}) leads
to a very high value of $z_f\sim 15$, which exceeds both
the age of dSph galaxies derived from observations of stars
(see, e.g., Weisz et al. 2014; Karachentsev et al., 2015)
and the redshift of reionization $z_{reio}\sim 9$ determined
by Planck (Ade 2016). Nonetheless the main significance
of the Lacey -- Cole model (\ref{rmod}) is the clear
introduction of the concept of the epoch (or redshift $z_f$)
of halo formation, what allows us to quantify correlation of
halo parameters with the process of growth of perturbations
and halo formation.

Evidently the basic assumptions of the simple model (\ref{rmod})
are not realistic. Indeed the real halos are formed
successively beginning from the core, what increases
$\langle\rho_{vir}\rangle$ as compared with (\ref{rmod}).
Further on the processes of merging and anisotropic matter
accretion unpredictably changes the density. Analysis of high
resolution simulations (Demia\'nski et al. 2011) shows that
influence of these random factors together with the strong
anisotropy of the early period of DM halo formation depends
upon the halo mass and, for example, it is moderate for
clusters of galaxies, within which, earlier formed galaxies
are observed as separate very dense elements. It is accepted
that for the simple $\Lambda$CDM cosmological model and for
more massive objects the expression (\ref{rmod}) describes
reasonably well both the observations and simulations. However,
as is discussed in section 3, the formation of low mass
objects is regulated by other factors and thus their structure
cannot be described by the same relations. These
comments attempt to explain the main reasons why the model
(\ref{rmod},\,\ref{mvir}) has limited applicability.

In order to describe the complex process of DM halo formation
in a wide range of virial masses of objects we use the more
general {\it phenomenological} relation
\be
\langle\rho_{vir}\rangle=18\pi^2\Phi(M_{vir})\langle
\rho_m(z_f)\rangle\approx\rho_{200}\Phi(M_{vir})(1+z_f)^3\,,
\label{pmod}
\ee
where $\Phi(M_{vir})\geq 1$ is a smooth slowly varying
function of $M_{vir}$. This relation preserves the
universality of (\ref{rmod}) and provides an unified
self consistent description of observed properties of DM
halos in a wide range of masses $10^6\leq M_{vir}/M_\odot\leq
10^{15}$\,. In particular, it reproduces the observed weak
mass dependence of the redshift $z_f$ and the density
$\langle\rho_{vir}\rangle$ for both clusters of galaxies
and low mass THINGs, LSB, UDG and dSph galaxies. Further
discussion of these problems can be found below in sections
4, and 5\,.

The expression (\ref{pmod}) allows us to take into account
anisotropy of the collapse, what decelerates the process of
objects formation and decreases both the virial density and
$z_f$. It can be used in more complex cosmological models
(such as, e.g., Nesseris \& Sapone, 2015).

Precise  observations of DM periphery of both clusters and galaxies are problematic  owing to the strongly irregular
matter distribution in their outer regions.  Hence, we use only estimates of the more stable parameter - the virial
mass of object as its leading characteristic. For example, sometimes the relation (\ref{rmod}) is used for description
of clusters of galaxies under the arbitrary assumption that the cluster is formed at the observed redshift,
$1+z_f\equiv 1+z_{obs}$. In this case the relation (\ref{rmod}) allows to determine formally the mean virial density
$\langle\rho_{vir}\rangle$ and, for a given mass of cluster $M_{vir}$, its virial radius, $R_{vir}$. However, this
result is evidently incorrect because, in fact, we can only conclude that $z_f\geq z_{obs}$, which is trivial for
galaxies. For clusters the difference between $1+z_f$ and $1+z_{obs}$ can be as large as $\sim 1.5 - 2$.

More stable and refined method to determine the redshift
of halo formation uses characteristics of the halo core
rather than its periphery. In this case  we use the
following expression for the concentration
\be
c(M_{vir},z_f)\approx 0.12M_{12}^{1/6}(1+z_f)^{7/3},\quad
M_{12}=\frac{M_{vir}}{10^{12}M_\odot},
\label{cmz}
\ee
(Demia\'nski\,\&\,Doroshkevich 2014; Ludlow et al. 2016). Together
with relations (\ref{nfw-ms}) and (\ref{pmod}) we get for the mean
density of the central core
\be
\langle\rho_s\rangle=\rho_{cc}M_{12}^{1/2}(1+z_f)^{10}\Phi(M_{vir})\,,
\label{cdns}
\ee
\[
\rho_{cc}=0.38\cdot 10^{-3}\frac{\rho_{200}}{f_m(c)}
\approx\frac{2.5}{f_m(c)}\frac{M_\odot}{kpc^3}=
\frac{1.9}{f_m(c)}\cdot 10^{-31}\frac{g}{cm^3}\,.
\]
This relation links the redshift $z_f$ with the virial mass
$M_{vir}$ and the mean density of halo core $\langle \rho_s
\rangle$ and allows to determine $z_f$. Application of this
approach requires additional observations. However, it is less
sensitive to random deviations of characteristics of periphery
of objects. Moreover for DM dominated high density objects
observed at $z_{obs}\ll 1$ (such as the dSph galaxies)
determination of the redshift $z_f$ through the parameters of
the central core is also preferred.

The main weakness of this approach is the possible impact of
baryonic component that can be specially important for
galaxies with moderate DM domination. However, for dSph galaxies
this effect can be comparable with the uncertainties in
measurements of parameters, what is clearly seen, when one
compares the results presented in Walker et al. 2009 and
Kirby et al. 2014 (see sections 5.5\,\&\,5.6).

Comparison of (\ref{pmod}) and (\ref{cdns}) indicates
that the density of halo core $\langle\rho_s\rangle$
is more sensitive than $ \langle\rho_{vir}\rangle $ to both
the virial mass and the redshift of
formation. However, for all observed samples (section 5) there
is a correlation between $z_f$ and $M_{vir}$:
 \be
\eta_f=(1+z_f)M_{12}^{0.077}\approx 4.1(1\pm 0.1)\,,
 \label{zm}
\ee
(see also Demia\'nski\,\&\,Doroshkevich 2014). Allowing
for this correlation we see from (\ref{cdns}) that actually
$\langle\rho_s\rangle$ is a weak function of both the virial
mass $M_{vir}$ and the redshift of formation $z_f$.

\subsection{Excursion set approach and shape of the power
spectrum}

It is seen from (\ref{pmod}) that the redshift of halo
formation $z_f$ is uniquely related to the mean DM density of
the virialized object. As it is demonstrated below sometimes
it is more convenient to use this redshift for description
of halos. In particular, it is well known that at redshifts
$z\geq 3$ the formation of galactic scale halos dominates,
but the typical mass of halos increases with time and massive
clusters of galaxies are mostly formed later at redshifts
$z\leq 2$. This correlation between the redshift of halo
formation $z_f$ and the halo mass $M_{vir}$ is described by
current models of halo formation.

Majority of such models (Press \& Schechter 1974; Bardeen et
al. 1986; Bond et al. 1991; Sheth \& Tormen 2002; 2004)
are based on the excursion set approach applied to the
initially Gaussian random density perturbations. They reduce
description of characteristics of the formed halos to the
problem of crossing of an appropriate barrier by particles
undergoing Brownian motion. These problems were discussed
very actively at the end of previous century with the use of
2D and 3D simulations with different power spectra and a
close link between the power spectrum and properties of DM
halos and Large Scale Structure had been demonstrated. The
short review of pertinent publications can be found in Sheth
\& Tormen 2002 \& 2004.

Here we apply this approach to the analysis of observed DM dominated objects in a wide range of virial masses
($10^6\leq M_{vir}/M_\odot\leq 10^{15}$), whicb allows us to demonstrate unexpectedly complex shape of the function
$z_f(M_{vir})$. Possible explanations of this fact are discussed in section 5.

All theoretical models of halos formation predict that the
distribution functions of halo characteristics are dominated
by a typical Gaussian term:
\be
dP(M_{vir})\propto \exp[-\alpha\Psi^2(M_{vir})]dM_{vir}\,,
\label{sz}
\ee
\[
\Psi(M_{vir})=\sigma_m(M_{vir})B(z_f(M_{vir}))\,,
\]
where $B(z_f)$ (\ref{Bz}, \ref{bzz}) describes
the growth of density perturbations and $\sigma_m$
is the dispersion of density perturbations and it is given by
\be
\sigma_m^2(M)=\frac{1}{2\pi}\int_0^\infty k^2p(k)W^2(k,M) dk\,.
\label{sigm}
\ee

Here $p(k)$ is the power spectrum and
\[
W(x)=3(sin\,x/x^3-cos\,x/x^2),\quad x=kr\propto kM^{1/3}\,,
\]
is the Fourier transform of the real-space top-hat filter
corresponding to a spherical mass $M$. Thus, the function
$\Psi(M_{vir})$ characterizes the amplitude of perturbations
with mass $M_{vir}$.

The usually used condition
\be
\Psi(M_{vir})=\sigma_m(M_{vir})B(z_f(M_{vir}))\approx const\,,
\label{const}
\ee
provides the expected approximate self similarity of
the process of halo formation and progressive growth with
time of the virial mass of halos. This model is consistent
with the main results of numerous 2D and 3D numerical
simulations performed with different box sizes, power
spectra and resolution (see, e.g., discussion in Sheth
\& Tormen 2002).

The coefficient $\alpha$ in (\ref{sz}) depends upon the
used normalization of $\sigma_m$ and specifies the universal
barrier that discriminates the linear from the nonlinear
evolution of a DM halo with the mass $M_{vir}$.  For example,
for the CDM -- like power spectrum and for the standard
normalization of perturbations on $\sigma_8$ the function
$\sigma_m$ and $\alpha$ are well fitted by the following
expressions
\be
\sigma_m=\frac{3.31\sigma_8 M_{12}^{-0.077}}{1+0.177M_{12}^{0.133}+
0.16M_{12}^{0.333}},\quad \alpha=\frac{1.686^2}{2\sqrt{2}}
\approx 1\,,
\label{sig_cdm}
\ee
where $M_{vir}= M_{12}10^{12}M_\odot$, 1.686 is the critical
overdensity (height of the barrier) and $\sigma_m\approx
\sigma_8$ for
\[
M_{12}=\frac{4\pi}{3}\frac{\langle\rho_m\rangle}{10^{12}M_\odot}
\left(\frac{8Mpc}{h}\right)^3\approx 210
\Theta_m\left(\frac{0.7}{h}\right)^3\,.
\]
Below in section 5 other normalizations of the functions
$\Psi(M_{vir})$ and $\sigma_m$ based on the observed parameters
will be used.

It is interesting that for $M_{12}\leq 1$ we get from
(\ref{bzz}), (\ref{zm}), and (\ref{sig_cdm}) that
\be
\Psi(M_{vir})\approx \frac{4.47\sigma_8}{(1+z_f)M_{12}^{0.077}}=
1.1\sigma_8\left(\frac{4.1}{\eta_f}\right)\approx const\,,
\label{par3}
\ee
which is consistent with (\ref{const}) and
demonstrates that properties of low mass objects are
in agreement with the CDM--like shape of the small scale
power spectrum. We will discuss in more details the
observed properties of the function $\Psi(M_{vir})$ in
section 5.

\section{The entropy of the relaxed DM halos}

Entropy is a very important characteristic of virialized DM
halos as it is conserved during their adiabatic evolution. The
entropy of a DM halo includes a component related to
small scale random perturbations of the compressed matter
and a component generated in the course of  violent
relaxation of the compressed matter. Evidently the former
depends directly upon the initial power spectrum while
the latter depends mainly upon the halo mass and the
period of halo formation.

Simulations demonstrate that indeed in accordance with the
Zel'dovich approximation during the early stages of halo
formation the very complex anisotropic matter compression
takes place and theoretical description of this period is
problematic (see, e.g., detailed discussion in Demia\'nski
et al. 2011). Published discussions of the process of
violent relaxation (Filmore and Goldrich 1987; Gurevich
and Zibin 1995; Lithwick and Dalal 2011; and others) use
the simplest spherical models and can illustrate, but
not reproduce, the real complex process of violent
relaxation.

Estimates of these components of entropy are given in
Appendix A. They can be compared with estimates obtained
from the analysis of observed objects.

The model (\ref{rmod},\,\ref{mvir}) describes the DM halo formation from a spherical cloud of DM particles at rest,
which implies small initial entropy of the cloud. Thus, this model demonstrates that some entropy is really generated
within the collapsed and relaxed DM halos even without any initial entropy. This entropy can be roughly estimated as
(\ref{asmvir}) \be \langle S_{vir}(M)\rangle\approx 4.3M_{12}^{0.74}cm^2keV \frac{\mu_{DM}^{2/3}}{\Phi^{1/3}}
\frac{\eta_f}{4.1} \frac{72}{\theta_{vir}}\,, \label{smvir} \ee where $\mu_{DM}=m_{DM}/m_b$, $m_{DM}\,\&\,m_b$ are the
masses of DM particles and baryons, $\theta_{vir}$ is the standard ratio of the gravitational and internal energy of
virialized objects averaged over 180 observed clusters with masses $10^{13}\leq M_{vir}/M_\odot\leq 10^{15}$: \be
\theta_{vir}=\frac{M_{vir}}{10^{12}M_\odot}\frac{1 Mpc}{R_{vir}} \frac{1 kev}{T_x}\approx 72(1\pm 0.08)\,, \label{MRT}
\ee
 and $T_x$ is the measured x-ray  halo temperature.
This entropy is comparable with observational estimates
(\ref{ffit}). However, these estimates apply mainly to
peripheries of halos with large entropy.

The second component includes contribution of random motions
linked with the random density perturbations. Using the
correlation between the redshift $z_f$ and the virial
mass of halo (\ref{zm}) we get for this component
(\ref{as2i})
\be
S_{r}(M_{vir})\approx 6.4M_{12}^{0.81}\mu^{5/3}g_{r}(\eta_f/4.1)^3
cm^2keV\,,
\label{s2i}
\ee
where $g_{r}$ is a factor that determines the fraction of
random energy accumulated by the halo. This value is also
comparable with the observational estimates (\ref{ffit}).

It can be expected that this channel of entropy generation
is more important for less massive halos.

\section{Expected properties of the relaxed DM halos}

We consider the DM halos as a one parametric sequence of
objects all properties of which depend upon their virial mass.
This means that we consider all DM halos as similar ones.
Together with the virial characteristics of halos, namely,
the mass, $M_{vir}$, radius $R_{vir}$, and density
$\langle\rho_{vir}\rangle$ we also consider the mean
characteristics of their central cores, namely, the density
$\langle\rho_s\rangle$, pressure, $\langle P_s\rangle$,
temperature $\langle T_s\rangle$, entropy $\langle S_s\rangle$
and DM surface density, $\langle\Sigma_s\rangle$. The very
important characteristic of a halo is the redshift of its
formation, $z_f$, that was introduced by (\ref{pmod}), it
approximately characterizes the end of the period of halo
formation and relaxation.

The DM temperature and velocity dispersion $\sigma_v$ in the
core usually are not observed, but within relaxed DM cores
$\sigma_v$ is close to the circular velocity, $v_c(r)$,
(Demia\'nski\,\&\,Doroshkevich 2014)
\be
\sigma_v^2(r)\approx v_c^2(r)\sqrt{r_s/r}\,.
\label{tc}
\ee
Because of this we can estimate the expected mean parameters
of a core of a DM halo. Expressions (\ref{nfw-m}) and
(\ref{cdns}) can be rewritten as follows:
\[
M(r_s)\approx 0.2M_{vir}\approx 0.2 M_{12}\cdot 10^{12}M_\odot,\quad
r_s\approx 3Mpc M_{12}^{1/3}(1+z_f)^{10/3}[\Phi(M_{vir})/f_m]^{1/3}
\]
With these relations we get for the mass dependence of the
parameters of core: 
\[
\langle T_s\rangle\approx \frac{m_{DM}\sigma_v^2}{2}\approx
\frac{m_{DM}}{2}\frac{GM(r_s)}{r_s}\approx
1.8eVM_{12}^{5/6}(1+z_f)^{10/3}(\Phi/f_m)^{1/3}\mu_{DM}\,,
\]
\be
\langle P_s\rangle\approx \langle n_{DM}(r_s)T_s\rangle\approx
10^{-7}eV/cm^3M_{12}^{4/3}(1+z_f)^{40/3}(\Phi/f_m)^{4/3}\,,
\label{ps}
\ee
\[
\langle S_s\rangle\approx \langle P_s/n_{DM}^{5/3}(r_s)\rangle
\approx 80 cm^2keV \frac{M_{12}^{1/2}}{(1+z_f)^{10/3}}
\mu_{DM}^{5/3}(f_m/\Phi)^{1/3}\,.
\]

Using the correlation between $z_f$ and $M_{vir}$ (\ref{zm})
we finally get
\[
\langle T_s\rangle\approx 0.2keVM_{12}^{0.6}\left(\frac{
\eta_f}{4.1}\right)^{10/3}\left(\frac{\Phi}{f_m}
\right)^{1/3}\mu_{DM}\,,
\]
\[
\langle\rho_s\rangle\approx \eta_\rho M_{12}^{-0.2}
\frac{\Phi}{f_m}\left(\frac{\eta_f}{4.1}\right)^{10}\,,
\]
\be
\langle P_s\rangle\approx \eta_p M_{12}^{0.4}\left(
\frac{\Phi}{f_m}\right)^{4/3}\left(\frac{\eta_f}{4.1}
\right)^{40/3}\,,
\label{psm}
\ee
\[
\langle S_s\rangle\approx \eta_s M_{12}^{0.73}\left(\frac{\eta_f}{
4.1}\right)^{-10/3}\left(\frac{f_m}{\Phi}
\right)^{1/3}\mu_{DM}^,\,.
\]
  where expected values of the constants are
\be
\eta_\rho\approx 2.6\cdot 10^6 \frac{M_\odot}{kpc^3},\quad
\eta_p\approx 22\frac{eV}{cm^3},\quad \eta_s\approx 0.9cm^2keV\,.
\label{e-etas}
\ee
These constants are constructed from parameters of the NFW
model (\ref{nfw-m}), the characteristic density $\rho_{cc}$
(\ref{cdns}) and the correlation between the virial mass and
the redshift $z_f$ (\ref{zm}).

The possibility to describe the internal structure of DM cores
in a wide range of masses by simple functions of the virial mass
only manifests the expected self similarity of the structure
of these objects.  For the observed objects values of these
parameters are obtained in
the next section and are presented in Table 1. For the best
sample of 19 dSph galaxies with $M_{vir}\leq 10^9M_\odot$ and 19
CLASH clusters with $M_{vir}\geq 10^{14}M_\odot$ we have
\be
\eta_\rho\approx 1.3\cdot 10^6(1\pm 0.9)\frac{M_\odot}{kpc^3}, \quad
\eta_s\approx 1.2(1\pm 0.5)cm^2keV,\quad
\eta_f\approx 4.1(1\pm 0.1)\,,
\label{s38}
\ee
which is quite similar to expectations (\ref{e-etas}).  Here
and below we use the correction factor
\be
\Phi(M_{vir})=(1+M_f/M_{vir})^{0.3},\quad M_f\approx 8\cdot
10^{12}M_\odot\,,
\label{phi_m}
\ee
that allows us to obtain an unified self consistent description
of both discussed galaxies and
clusters of galaxies. Thus, for larger masses $M_{vir}\gg M_f$,
$\Phi\rightarrow 1$, and expressions (\ref{rmod}) and
(\ref{pmod}) become identical to each other. On the other hand,
in the opposite case $M_{vir}\ll M_f$ the function
(\ref{phi_m}) allows to reconcile  theoretical expectations
(\ref{psm}) with observations (\ref{ffit}) presented in
section 5. For low mass dSph galaxies this approach decreases
the redshift of formation $z_f$ down to values consistent
with the observed age of stars and Planck estimates of the
redshift of reionization.

Of course the function $\Phi(M_{vir})$ (\ref{phi_m}) should be
considered only as the first approximation. But to
obtain more refined and justified description of the redshift
of formation $z_f$ (\ref{pmod}) and DM halo parameters
(\ref{psm}) we need to have both richer observational data with
only moderate scatter and corresponding high resolution
simulations.

As is seen from (\ref{zm}) and (\ref{cmz}) the so defined
concentration is only weakly depended on the virial mass,
\be
\langle c\rangle\approx 3(\eta_f/4.1)^{7/3}M_{12}^{0.004}\,,
\label{cc}
\ee
which is confirmed -- in the range of data uncertainties --
by the results presented in Table 1. It is interesting that
from these relations immediately follows the widely
discussed weak mass dependence of the DM surface density
of both the core, $\Sigma_s$, and the virialized object,
$\Sigma_{vir}$. Indeed,
\[
\langle\rho_{vir}\rangle = 64\rho_{200}
\frac{\Phi(M_{vir})}{M_{12}^{0.21}}\left(\frac{\eta_f}{4.1}
\right)^3\approx \frac{45\cdot 10^{-5}}{M_{12}^{0.21}}
\left(\frac{\eta_f}{4.1}\right)^3\Phi\frac{M_\odot}{pc^3}\,,
\]
\[
\Sigma_{vir}=\langle\rho_{vir} R_{vir}\rangle\approx
35M_{12}^{0.20}\left(\frac{\eta_f}{4.1}\right)^2
\Phi^{2/3}\frac{M_\odot}{pc^2}\,,
\]\be
\Sigma_s=\langle\rho_s r_s\rangle=\frac{\Sigma_{vir}c^2}{5 f_m(c)}
\approx 65M_{12}^{0.28}\left(\frac{\eta_f}{4.1}
\right)^{6.7}\Phi^{2/3}\frac{M_\odot}{pc^2}\,.
\label{svir}
\ee
Such stability of  both surface densities $\Sigma_s$
and $\Sigma_{vir}$ was discussed in many papers (see, e.g.,
Spano et al., 2008; Donato et al. 2009; Salucci et al.
2011; Demia\'nski \& Doroshkevich 2014; Saburova \& Del
Popolo, 2014). Estimates of $\Sigma_s$ (\ref{svir}) and that
listed in Table 1 are similar to results of Kormendy \&
Freeman 2015. They support the expected self similarity of
the process of DM halo formation and the proposed description
of the halo structure. In particular it supports the proposed
modification of the model (\ref{pmod}).

\section{Observed characteristics of DM dominated galaxies and
clusters of galaxies}

For our analysis we used more or less reliable observational
data for $\sim 19$ DM dominated clusters of galaxies, 30
groups of galaxies, one UDG, $\sim 30$ dSph, and $\sim 11$
THINGS and
LSB galaxies. The analysis of observations of the dSph
galaxies in the framework of our model was performed
in Demia\'nski \& Doroshkevich 2014. Here we improve this
analysis and compare the observational results obtained
within a wide interval of virial masses with theoretical
expectations.

Now there are more or less reliable observational data for
at least $\sim 300$ clusters of galaxies (Pointecouteau et
al. 2005; Arnaud et al., 2005; Pratt et al., 2006; Zhang et
al., 2006; Branchesi et al., 2007; Vikhlinin et al., 2009;
Pratt et al. 2010; Suhada et al. 2012; Moughan et al. 2012;
Bhattacharya et al. 2013; Merten et al. 2015). However,
usually the central cluster characteristics are not directly
observed and are obtained by a rather complex procedure
(see, e.g., Bryan\,\&\,Norman 1998; Pointecouteau et al. 2005;
Vikhlinin et al. 2009; Lloyd--Davies et al. 2011;
McDonald et al., 2013). In spite of the rapid progress of
investigations of clusters of galaxies, recent publications
discuss mostly the general cluster characteristics such as
their observed redshift $z_{obs}$, virial mass, $M_{vir}$,
radius, $R_{vir}$, and average X-ray temperature, $T_x$. However,
in this paper we are mainly interested in discussion
of more stable central regions of clusters and, in particular,
in concentrations and the central pressure and entropy
of DM component. Unfortunately such data are very limited
and for our analysis we can use only the CLASH survey. For
comparison and in order to demonstrate the possible complex
impact of baryonic component we also consider 83 SPT
clusters (McDonald et al. 2013).

In this section we consider properties of the central cores
of virialized DM halos using the approximation
summarized in the previous section. We characterize
halos by their virial mass $M_{vir}$ and redshift of formation,
$z_{f}$. We assume that at $z\leq z_{f}$ the halos
mass and core temperature and density do not change
significantly.

\subsection{The CLASH clusters}

For 19 clusters of the CLASH sample (Merten et al. 2015) in
addition to the usually presented observed virial mass of
clusters, $M_{vir}$, there are also estimates of the size
and density of central core, $r_s,\,\&\,\rho_0$. For this
survey the published parameters $R_{vir},\,\&\,\rho_{vir}$ are
obtained under the assumption that $z_f=z_{obs}$ and they are
not used in our analysis. The virial mass of
clusters, $M_{vir}$, presented in Merten et al. 2015 is
close to the estimates of $M_{vir}$ obtained by Umetsu et
al. 2014, what confirms reliability of these estimates.

It is important that for these clusters parameters of
cores can be used without serious corrections and the
concentrations $\langle c\rangle\leq 4$ are quite moderate.
The similarity of $r_s$ and $r_{vir}$ demonstrates a limited
impact
of central galaxies and/or cooled baryonic component for
these clusters (see, e.g. Mantz 2014) and allows to consider
these data as typical among DM dominated clusters. Main
results are presented in Table 1 and are plotted in Figs.
\ref{fprs} and \ref{sig6}.

The difference between our estimates of $z_f$ and $z_{obs}$
for this survey
\be
\langle (1+z_f)/(1+z_{obs})\rangle\approx 1.7(1\pm 0.2)\,,
\label{zfzobs}
\ee
demonstrates arbitrary character of the assumption that
$z_f\equiv z_{obs}$ and conventional character of results
based on this assumption.

Using (\ref{MRT}) we get for this survey
\be
\langle\theta_{vir}\rangle=77(1\pm 0.2),\quad
\langle M_{12}\rangle=900(1\pm 0.3)\,,
\label{thet-20}
\ee
\[
\langle \rho_s\rangle=5\cdot 10^5(1\pm 0.5)M_\odot/kpc^3,
\quad \langle 1+z_f\rangle=2.3(1\pm 0.1)\,,
\]
 and for the $\Lambda$CDM spectrum (\ref{sig_cdm})
\be
\langle \Psi(M)\rangle=0.4\sigma_8(1\pm 0.1)\,.
\label{par1}
\ee

\subsection{The SPT -- clusters}
For 83 clusters selected by the South Pole Telescope (Reinhardt 
et al. 2013; McDonald et al. 2013; Ruel et al. 2013;
Saliwanchik et al. 2015) the central baryonic density, 
temperature and entropy are given at radius $r\leq 0.012
R_{500}$ while the expected radius of the cluster core is 
$r_s\sim (0.15-0.25)R_{500}$. For these clusters also the
standard virial masses $M_{500}$ are known, but no information 
on the DM component is provided, which prevents us to
accurately reconstruct  parameters of these cores. However, 
these clusters nicely illustrate the complex behavior of
their baryonic cores.

This sample is clearly divided into two groups with
different properties of their observed baryonic component.
One of them contains
35 clusters with baryonic number density $\langle n_b
\rangle\geq 8\cdot 10^{-3} cm^{-3}$ and entropy of the central
core $\langle S_b\rangle\leq 90 cm^2keV$,  what indicates the
noticeable cooling of the baryonic component. Owing to the
thermal instability this process results in the formation of
two component medium -- cold low mass high density
baryonic subclouds within hot low density diffuse baryonic gas
(Doroshkevich \& Zel'dovich 1981). In these clusters the
observed density relates to the denser fraction while the
observed temperature relates to the diffuse hot gas and
random velocities of clouds (see, e.g., Khedekar et al. 2013;
Battaglia et al., 2015; Adam et al., 2015). This means
that for these clusters the real central pressure and entropy
of the hot baryonic component are close to that measured
for the hotter subsample of clusters.

The hotter group contains 38 clusters with baryonic entropy
\[
\langle S_b\rangle = 270(1\pm 0.5)cm^2keV\,.
\]
Unfortunately we have no information about the central
characteristics of their DM component.  However, if we assume
that in central regions
\be
\langle\rho_{DM}\rangle\approx 3 \langle\rho_{b}\rangle\,,
\label{sptt}
\ee
then our estimates of cluster characteristics become close
to (\ref{thet-20}) and (\ref{par1}). This choice seems to
be quite reasonable.  Indeed for intercluster medium we expect
that $\langle\rho_{m}\rangle/\langle\rho_{b}\rangle \approx 6$,
see (\ref{basic}). In clusters the baryonic component tends to
cool and to settle in their cores, what decreases the
ratio $\langle\rho_{m}\rangle/ \langle\rho_{b} \rangle$. For
these 38 clusters with the accepted value of (\ref{sptt})
the mean results are quite similar to those obtained for the
CLASH survey, in particular,
\be
\langle
(1+z_f)\rangle=2.9(1\pm 0.1),\quad \langle \Psi(M)\rangle=0.34
\sigma_8(1\pm 0.1)\,.
\label{par11}
\ee

\subsection{The groups of galaxies}

30 suitable groups of galaxies with $z_{obs}\ll 1$,
 masses $1.2\leq M_{12}\leq 200$ and sizes $100kpc\leq
R\leq 700kpc$ are taken from the catalog of Makarov \&
Karachentsev 2011. For these objects we cannot separate
the central core and  estimate its parameters. But these
data allow us to estimate -- with large scatter -- the
mean density and the redshift of formation as
\[
\langle M_{12}\rangle=40(1\pm 1),\quad
\langle\rho_{vir}\rangle=2.7\cdot 10^5(1\pm 0.9)M_\odot/kpc^3,
\]
\be
\langle 1+z_f\rangle=3.1(1\pm 0.3),\quad
\langle \Psi(M)\rangle=0.74\sigma_8(1\pm 0.3)\,,
\label{grup1}
\ee
and so to partly fill the empty region
in Fig. \ref{sig6} between the galaxies and clusters of
galaxies. Main results of our analysis are plotted
in Fig. \ref{sig6} and presented in Table 1.

\subsection{The THINGS and LSB galaxies}

For 11 LSB and THING galaxies (de Blok et al. 2008; Kuzio
de Naray et al. 2008; Chemin et al. 2011) the observed
rotation curves are measured up to large distances, what
allows us to estimate, with a reasonable reliability,
the virial masses $M_{vir}$ and the mean virial density,
$\langle\rho_{vir}\rangle$, and, finally, using the relation
(\ref{pmod}) to estimate the redshift of formation and the
virial mass of these objects, we get
\[
\langle M_{12}\rangle=0.24(1\pm 0.8),\quad
\langle\rho_{vir}\rangle=6.2\cdot 10^5(1\pm 0.9)M_\odot/kpc^3,
\]
\be
\langle 1+z_f\rangle=2.7(1\pm 0.1),\quad
\langle \Psi(M)\rangle=1.6\sigma_8(1\pm 0.2)\,.
\label{galac}
\ee

Non the less the complex internal structure of these galaxies
and the significant influence of stars, discs and diffuse
baryonic component restricts the number of objects for which
reasonable DM characteristics can be derived. For 11
galaxies   results of our analysis are presented in
Table 1 and in Figs. \ref{fprs} and \ref{sig6}. But even
for these galaxies the weak dependence of the measured
rotation curves upon their radius demonstrates that the
accepted virial mass of galaxies is noticeably underestimated
and reliability of obtained results is in question.

\subsection{The dSph galaxies}

Recently properties of the  dSph galaxies were discussed in
many papers. Thus, the main observed parameters of 28 dSph
galaxies are listed and discussed in Walker et al. 2009
\& 2011; Penarrubia et al. 2010; for 13 And galaxies with
similar properties results are listed in Collins et al. 2014;
Tollerud et al. 2012 \& 2014. Extensive list of similar
galaxies is also given by McConnachie 2012. Published
characteristics of these galaxies vary from paper to paper
and are presented with significant scatter. From samples
presented in Walker et al. 2009 \& 2011; Collins et al.
201); Tollerud et al. 2012 \& 2014 we selected 19
objects with high DM density, what suggests that these
objects were formed at high redshifts and can be considered
as samples of earlier galaxies responsible for reionization.
This means that for these galaxies the redshift of formation
$z_f$ should be comparable with redshift of reionization
$z_{reio}\sim 9$ determined by WMAP and Planck missions. For
other 22 objects of these surveys situation is not so clear
and either their published parameters are unreliable or
perhaps they were formed later as a surviving companions
of more massive objects.

Our sample includes objects in a wide range of masses, what
allows us to reveal more reliably the mass dependence of
their redshift of formation (Demia\'nski \,\&\, Doroshkevich
2014). But in this case we have to deal with parameters of
DM at the projected half--light radius and they must be
recalculated to the virial mass and core parameters, what
introduces additional uncertainties. In spite of these problems
it is very interesting to compare characteristics of
these galaxies with characteristics of clusters of galaxies
presented in the previous sections and with theoretical
expectations (\ref{psm}).

Comparison of the observed circular velocity $v_c$ and velocity
dispersion $\sigma_v$ shows that for the subpopulation of 19
denser objects the half--light radius $r_{1/2}$ corresponds to
concentration
\[
 v_c^2(r)/\sigma_v^2(r)\approx \sqrt{r_{1/2}/r_s}=\sqrt{c_{1/2}}\,,
\]
\be
\langle c_{1/2}\rangle=\langle r_{1/2}/r_s\rangle \approx
1.6(1\pm 0.04)\,,
\label{c_1/2}
\ee
what allows us to restore the main characteristics of the
sample. Thus, the virial mass of the galaxies can be
estimated with reasonable precision using (\ref{nfw-m}) as
\[
M_{vir}=M_sf_m(c)\approx M_s=M_{1/2}/f(c_{1/2})\,,
\]
because $0.7\leq f_m(c)\leq 1$, for concentrations of interest
$3.5\leq c\leq 5$. For $c\leq 4$ this approach
overestimates the virial mass by about 15 - 20\%, which is
comparable with the precision of measurements. For $4\leq
c\leq 5$ errors become still smaller. Expression (\ref{cmz})
can be used to test the internal agreement of such
approach. This sample accumulates objects in a wide range of
masses $10^6\leq M_{vir}/M_\odot\leq 10^{9}$, which had
been formed at high redshifts
\be
 12.8\geq 1+z_f\geq 8.6,\quad \langle 1+z_f\rangle\approx
9.9(1\pm 0.1)\,.
\label{zphi}
\ee
The basic parameters of these objects are found using the
function $\Phi(M_{vir})$ (\ref{phi_m}) in (\ref{pmod},\,
\ref{ps}, \ref{psm}) which, in contrast with (\ref{rmod},
\,\ref{mvir}), allows us to reconcile the theoretical
expectations (\ref{ps}, \ref{psm}) for parameters
of the DM dominated halos with the observed ones (\ref{ffit}).
This choice of $\Phi(M_{vir})$ also decreases the redshift
of formation of dSph galaxies down to (\ref{zphi}). Such
estimates of $z_f$ are consistent with results of Planck
$z_{reio}\sim 9$ (Ade et al. 2016) and are confirmed by
estimates of ages of stars $z_f\geq 5 - 6$ (see, e.g.,
Weisz et al. 2014; Karachentsev et al. 2015). Large
scatter of our estimates is mainly caused by
the large scatter of observational parameters. Basic
characteristics of these objects are presented in Table 1
and are discussed in section 6.

Using (\ref{MRT})  for the observed parameters of this survey
at the projected half--light radius $r_{1/2}$, we get
 \be
\langle r_{1/2}\rangle\approx 140pc,\,\, \langle M_{1/2}\rangle
\approx 2.4\cdot 10^7M_\odot,\,\,
\langle\theta_{1/2}\rangle\approx120\,.
\label{thet-dsph}
\ee
This value of $\langle\theta_{1/2}\rangle$ is larger than
(\ref{thet-20}) by a factor of 1.5\,. It can be partly caused
by our using of the X--ray temperature $T_x$ in
(\ref{thet-20}) and velocity dispersion of stars in
(\ref{thet-dsph}).

However, this difference can indicate real differences in the
internal structures of dSph galaxies and clusters of
galaxies, which can be induced by different factors that
determine the entropy generation in these objects. As was
discussed in section 3 for massive clusters of galaxies the
entropy of DM component is formed mainly by the violent
relaxation of compressed matter, while for dSph galaxies
significant contribution comes from initial random motions of
matter.

For the $\Lambda$CDM spectrum (\ref{sig_cdm}) and
redshift (\ref{zphi}) we get for the dSph data
\be
\langle \Psi(M)\rangle=1.1\sigma_8(1\pm 0.08)\,,
\label{par2}
\ee
which is quite similar to (\ref{par3}).

These estimates of $\langle\Psi(M)\rangle$ are by a
factor of 2.75 larger than those  obtained for the CLASH
survey (\ref{par1}). This difference suggests the
corresponding difference in the amplitude of power
spectrum at large and small scales. We will discuss this
problem in more details in section 7.1 .

Our results presented in this section depend upon the choice
of the function $\Phi$ (\ref{phi_m}). To demonstrate the
real influence of this factor we reanalyzed the same observational
data in the same manner assuming that
\be
\Phi(M_{vir})=1\,. \label{phi_1}
\ee
Naturally the observed physical characteristics of objects,
$\langle\rho_s\rangle$, $\langle p_s\rangle$, and $\langle S_s
\rangle$ remain the same as listed in Table 1 and
plotted in Fig. \ref{fprs}, but now differences between
expectations (\ref{mvir},\,\ref{psm}) and observations
(\ref{ffit}) become significant. Moreover such choice strongly
increases the redshifts of formation and concentrations
of halos up to
\be
\langle 1+z_f\rangle\approx 15.7(1\pm 0.2),\quad
\langle\eta_f\rangle\approx 6.5(1\pm 0.07)\quad
\langle c\rangle\approx 11(1\pm 0.2)\,.
\label{zflg}
\ee
In turn this immediately changes all self similar parameters
and now we get
\be
\langle\eta_\rho\rangle\approx 220 M_\odot/kpc^3,\quad
\langle\eta_p\rangle\approx 6\cdot 10^3
eV/cm^3,\quad \langle\eta_s\rangle\approx 0.2 cm^2eV\,,
\label{phi=1}
\ee
instead of corresponding parameters listed in
Table 1\,. This growth of $z_f$ decreases estimate (\ref{par2})
of the function $\langle \Psi(M)\rangle$ down to
\be
\langle\Psi(M)\rangle=0.73\sigma_8(1\pm 0.2)\,,
\label{par4}
\ee
but as before this estimate significantly exceeds the
value (\ref{par1}) obtained for the CLASH survey.

Thus, the choice $\Phi=1$ destroys the agreement between
expectations (\ref{psm}) and observations (\ref{ffit}) and
removes the attractive inference about the self similarity
of virialized DM dominated objects in a wide range of masses.
On the other hand these values of $\langle z_f\rangle$
and $\langle c\rangle$ (\ref{zflg}) appear unreliably large.
These estimates can be directly tested with more detailed
observations even with the already known dSph objects.
Nonetheless the latest observations of PLANCK decreases
the redshift of reionization from $z_{rei}\sim 10.5$
(Komatsu et al. 2011) down to $z_{rei} \sim 8.9$  (Ade
et al. 2016; Robertson et al. 2015; Mitra et al. 2015),
which is quite similar to our estimates (\ref{zphi}).
The physical reason to use the expression (\ref{pmod}) with
the function $\Phi(M_{vir})$ as (\ref{phi_m}) were discussed
in sections 2.2 \& 3.

\subsection{The Kirby catalog of dSph galaxies}

Second catalog of dSph galaxies (Kirby et al. 2014) contains
30 satellites of Milky Way and Andromeda with independent
measurements of their parameters. Thus, some of them
significantly differ from those presented in Walker et al.
2009 \& 2011 and Tollerud et al. 2012 \& 2014. In
particular instead of (\ref{thet-dsph}) we have for
this survey at the projected half--light radius $r_{1/2}$
\be
\langle r_{1/2}\rangle=730pc,\quad \langle
M_{1/2}\rangle\approx 0.7\cdot 10^8\,M_\odot,\quad
\langle\theta_{1/2}\rangle=150\,,
\label{thet-krb}
\ee
with significant increase of $\langle r_{1/2}\rangle$ and
$\langle M_{vir}\rangle$. On the other hand, for this sample
we get
\be
\langle 1+z_f\rangle=8.1(1\pm 0.1),\quad \eta_f\approx 4.1(1\pm
0.1),\quad \langle \Psi(M_{vir})\rangle=1.2\sigma_8(1\pm 0.1)\,,
\label{par_kir}
\ee
which is quite similar to the previous results (\ref{par3},
\ref{zphi}, \ref{par2}). Other results for this survey are
presented in Table 1 and are plotted in Figs 1 \& 2.

This comparison of results obtained for two sets of
observations of the same objects allows to verify
representativity and reliability of obtained estimates.
These data were analyzed in the same manner as the previous
ones and the main results are close to those obtained for
the previous sample.

\subsection{Ultra diffuse galaxies}

Now there are new possibilities of observations of the Ultra
Diffuse Galaxies (UDG) near M101 (Merrit et al. 2014),
in the Virgo cluster (Liu et al. 2015; Beasley et al. 2016),
in the Coma cluster (van Dokkum et al. 2015; 2016), and in
the Pisces-Perseus supercluster (Martinez-Delgado et al. 2016).
For one of this object -- DM dominated galaxy Dragonfly
44 (van Dokkum et al. 2016) -- there are observations of
$M_{1/2},\,\,r_{1/2},\,\&\,\sigma_v$\, that is, the same data as
for the dSph galaxies. With these data we can repeat the
analysis performed in two previous subsections. Thus, we get
for this galaxy
\[
M_{vir}\approx 1.4\cdot 10^{10}M_\odot,\quad \rho_s\approx
4.2\cdot 10^{8}M_\odot/kpc^3,\quad p_s\approx 7.8eV/cm^3,
\quad S_s\approx 0.078cm^2 keV\,,
\]
\be
\eta_p=3.4eV/cm^3,\quad \eta_\rho=2.4\cdot 10^7M_\odot/kpc^3,
\quad \eta_s=3.3cm^2keV,\quad c\approx 2.8\,,
\label{udg}
\ee
\[
1+z_f\approx 5.2,\quad B^{-1}(\eta_f)\approx 3.8,\quad
\eta_f=3.8,\quad \Psi=1.1\sigma_8\,.
\]
These data are presented in Figs. \ref{fprs}\&\,\ref{sig6}.

As is seen from these Figures this galaxy is located in the
halfway between the CLASH clusters and dSph galaxies and
quite well fits with other data plotted in these Figures. This
can be considered as the independent evidence in favor
of our approach and inferences. We hope that further
investigations of UDG galaxies will improve our
results.

\begin{table*}
\begin{minipage}{140mm}
\caption{Characteristics of DM halos}
\label{tbl1}
\begin{tabular}{lrr rrr} 
   &   &    &   &   &   \cr
\hline
    &  CLASH&Groups&Galaxies&dSph&Kirby\cr
\hline
$N_{obj}$  &19~~~&30~~~&11~~~&19~~~&30~~~\cr
$\langle M_{12}\rangle$&$\sim 870$~~~&$\sim 40$~~~&$0.2(1\pm 0.8)$
&$\sim 10^{-4}$&$\sim 10^{-4}$\cr
$\langle\rho_s\rangle$ &$5(1\pm 0.5)$&$2.7(1\pm 0.9)$&
$3.4(1\pm 0.9)$&$10^5(1\pm 0.9)$&$10^5(1\pm 0.9)$\cr
$\langle p_s\rangle$&$224(1\pm 0.7)$& &$0.2(1\pm 0.8)$&
$10^{-5}(1\pm 0.9)$&$10^{-4}(1\pm 0.9)$\cr
$\langle T_s\rangle$&$10(1\pm 0.3)$& &$0.1(1\pm 0.5)$&
$10^{-3}(1\pm 0.8)$&$10^{-3}(1\pm 0.8)$\cr
$\langle S_s\rangle$&$1.6(1\pm 0.9)$& &$0.1(1\pm 0.9)$&
$10^{-6}(1\pm 0.9)$&$10^{-5}(1\pm 0.9)$\cr
$\langle c\rangle \,$$        $&$3.7(1\pm 0.2)$& & &
$3.7(1\pm 0.2)$&$3.3(1\pm 0.2)$\cr
$\langle\Sigma_s\rangle$&$3.4(1\pm 0.3)$& & &
$1.0(1\pm 0.5)$ &$0.8(1\pm 0.5)$\cr
$\langle\eta_\rho\rangle$&$16(1\pm 0.5)$& &
$0.6(1\pm 0.9)$&$2.2(1\pm 0.7)$&$1.5(1\pm 0.8)$\cr
$\langle\eta_s\rangle$&$1.2(1\pm 0.1)$& &
$31.(1\pm 0.9)$&$0.9(1\pm 0.3)$&$1.1(1\pm 0.4)$\cr
$\langle\eta_f\rangle$         &$3.9(1\pm 0.1)$&
$3.9(1\pm 0.3)$&$2.4(1\pm 0.2)$&$4.2(1\pm 0.1)$&$4.1(1\pm 0.1)$\cr
$\langle 1+z_f\rangle$ \,         &$2.3(1\pm 0.1)$&
$3.1(1\pm 0.3)$&$2.3(1\pm 0.2)$&$9.9(1\pm 0.1)$&$8.1(1\pm 0.1)$\cr
$\langle\Psi\rangle/\sigma_8$ \,&$0.4(1\pm 0.1)$&
$0.7(1\pm 0.3)$&$1.6(1\pm 0.2)$&$1.1(1\pm 0.1)$&$1.1(1\pm 0.1)$\cr
\hline
\end{tabular}

Here $\langle M_{12}\rangle=\langle M_{vir}\rangle/10^{12}M_\odot$,
$\langle\rho_s\rangle$\,\&\,$\langle\eta_\rho\rangle$ are measured
in $10^5 M_\odot/kpc^3$, $\langle p_s\rangle$ in $eV/cm^3$,
$\langle T_s\rangle \,$ in keV, $\langle S_s\rangle$ in
$10^2cm^2keV$, $\langle\Sigma_s\rangle$ in $100 M_\odot/pc^2$,
$\langle\eta_s\rangle$ in $cm^2keV$,
\end{minipage}
\end{table*}

\subsection{Sample of 8 dSph galaxies}

Recently a very detailed discussion of 8 dSph
galaxies was performed by Burkert (2015). In this paper
application of more complex two component isothermal and
King's models are considered. His main results are quite similar
to our estimates. It is important that estimates of the
central densities exceed our values by a factor $\sim 2$,
what increases the corresponding redshifts $z_f$ by 20
-- 30\% . Further investigations of dSph galaxies are
clearly needed.

\section{Cores of DM dominated virialized objects}

\begin{figure}
\centering
\epsfxsize=7. cm
\epsfbox{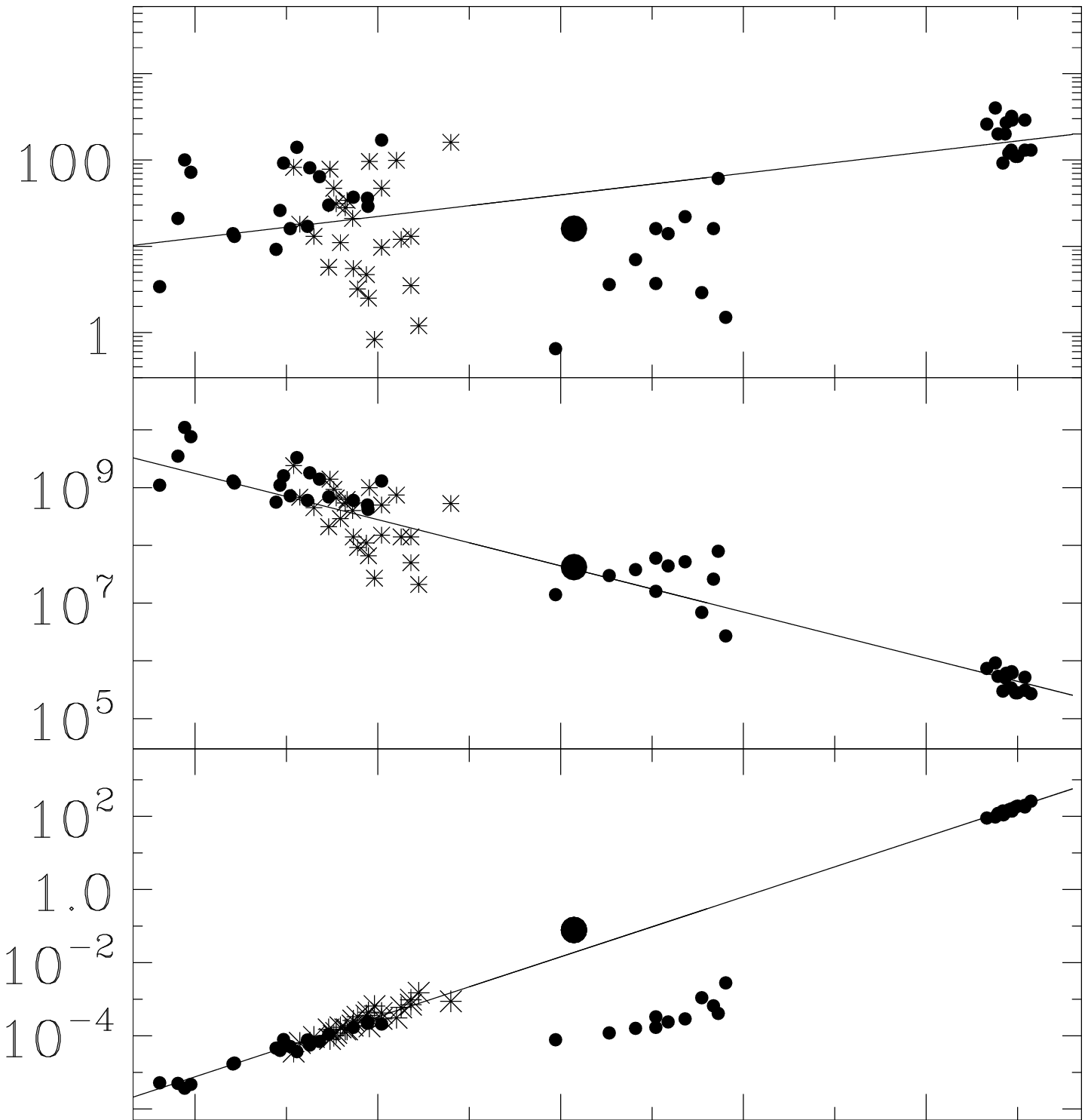}
\vspace{0.7cm}
\caption{The functions $\langle P_s(M_{vir}\rangle,\,\langle
\rho_s(M_{vir}\rangle),\,\&\,\langle S_s(M_{vir})\rangle$ are
plotted for the two samples of dSph galaxies (section 5.5, left
points and section 5.6, stars), THING and LSB galaxies (central
points) and CLASH clusters of galaxies (right points). For
the Dragonfly 44 values (\ref{udg}) are plotted by filled dark
circle. Fits (\ref{ffit}) are plotted by solid lines.
}
\label{fprs}
\end{figure}

\subsection{Basic characteristics of the DM cores}
For two samples of dSph galaxies, Dragonfly 44 galaxy, and CLASH
clusters of galaxies parameters of their DM cores (the
pressure, $P_s$, density, $\rho_s$, and entropy, $S_s$) are
plotted in Fig. \ref{fprs} and are fitted by the
expressions:
\[
\langle P_s\rangle=140(1\pm 0.8)eV/cm^3M_{12}^{0.1}\,,
\]
\be
\langle \rho_s\rangle=10^7(1\pm 0.6)\frac{M_\odot}{kpc^3}
M_{12}^{-0.45}\,,
\label{ffit}
\ee
\[
\langle S_s\rangle=0.66(1\pm 0.4)keV cm^2 M_{12}^{0.85}\,,
\]
\[
\langle 1+z_f\rangle=4.1(1\pm 0.17) M_{12}^{-0.077}\,,
\]
where again $M_{12}=M_{vir}/10^{12} M_\odot$. For comparison
in Fig. \ref{fprs} the same functions are also plotted
for THING and LSB galaxies. Large deviations of these
parameters from the fits are caused by uncertainties in estimates
of the influence of baryonic component, of the virial mass, and
other parameters of the DM cores. Large scatter of the
functions (\ref{ffit}) reflects mainly the large scatter of
the observational data and the natural random variations of
object characteristics.

Our analysis revealed some unexpected peculiarities in
characteristics of cores of the observed DM dominated virialized
objects. First is the small variation of parameters
$\eta_\rho,\,\eta_p,\,\&\, \eta_s$ presented in Table 1 for
the five samples of observed virialized objects in a wide range
of virial masses.  The large scatter of presented results
is partly caused by scatter of the observed parameters and by
probable influence of neglected factors of evolution such
as the impact of baryonic component. This scatter is small for
the entropy parameter $\eta_s$ as the entropy is the
more stable characteristic of objects. Moreover for the CLASH
sample all parameters of cores are quite close to the
expected ones (\ref{psm}). We consider this similarity as an
evidence in favor of the self similar character of the
internal structure of observed DM dominated objects.

Next is a weak variation of the DM pressure in cores of such
halos with the virial mass, redshift of formation and
other characteristics of these objects. Thus, for CLASH and dSph
samples the mean virial masses differ by seven orders
of magnitude while the central pressures differ by a factor
$\sim 20$ only. In contrast, the entropy of cluster cores
significantly  -- by a factor of $10^6 - 10^7$ exceeds the entropy
of DM dominated objects of galactic scale.

Other manifestation of this peculiarity is the weak
mass dependence of the surface density
of cores (see, e.g., Spano et al., 2008; Donato et al.
2009; Salucci et al. 2011; Demia\'nski \& Doroshkevich
2014; Saburova \& Del Popolo, 2014; Kormendy \& Freeman
2015). For the DM dominated objects in the CLASH
sample
\be
\langle\Sigma_s\rangle\approx 340(1\pm 0.3)\frac{M_\odot}{pc^2}\,,
\label{sig-rho}
\ee
and for the objects in dSph sample
\[
\langle\Sigma_s\rangle\approx 100(1\pm 0.5)
\frac{M_\odot}{pc^2}\,.
\]
These results are consistent with weak mass dependence
of the DM pressure (\ref{ffit}) as
\[
P_s\propto \rho_sM_s/r_s\propto \Sigma_s^2\,.
\]
This shows that (\ref{sig-rho}) is a natural result
for cores of relaxed DM halos formed from perturbations
with the CDM--like power spectrum.

In contrast with (\ref{mvir}) the quite moderate mass dependence
of the observed parameters (\ref{ffit},\,
\ref{sig-rho}) confirms the validity of expression (\ref{phi_m})
used for description of DM dominated relaxed halos.
Indeed,  for the function $\Phi$ as defined by (\ref{phi_m}) both
expressions (\ref{s38}) and (\ref{ffit}) become very
similar to each other and they fit the observational data for the
samples with similar scatter. Both expressions for
the entropy are almost identical, but the expression (\ref{ffit})
for $\langle P_s\rangle$ more clearly demonstrates
weak variations of the DM core pressure with the virial mass of
halos. Both of these fits are very preliminary and
should be essentially improved with richer observational data
with small scatter.

It is important that these basic characteristics of objects
depend upon the virial mass only while the earlier used
relations (sections 2\,\&\,3) relay on  explicit dependence upon
the redshift $z_f$. Thus, the relations (\ref{psm}) and
(\ref{ffit}) confirm that the redshift $z_f$ is also an
explicit function of the virial mass of objects (see also
discussion in Demia\'nski and Doroshkevich 2014), which is
consistent with the basic ideas of the present day theories
of galaxy formation (e.g. Press -- Schechter, 1974; Bardeen
et al. 1986; Bond et al. 1991; Sheth \& Tormen 2002,
2004). Moreover similarity of the power induced in
expressions (\ref{sig_cdm}) and (\ref{zm}) confirms the
CDM--like shape of the small scale power spectrum.

As well as properties of DM virialized halos discussed in
section 3 these inferences have only statistical significance.
Nonetheless allowing for the wide interval of virial masses
under consideration, weak variations of the parameters
$\eta_\rho,\,\eta_p,\,\&\,\eta_s$ introduced in (\ref{psm})
and presented in Table 1 demonstrate the self similarity of
basic properties of the observed cores of DM
dominated galaxies and clusters of galaxies.

Such self similarity naturally appeared in simulations with
the simplest power spectrum $p(k)\propto k^n$. For the CDM
model simulations show similarity of the dimensionless
characteristics of the DM halos such as the NFW density and
pressure profiles. On the other hand the regular character
of the CDM initial power spectrum (Bardeen et al. 1986) sets
up correlations between the virial masses of DM objects and
their redshifts of formation. For the observed DM dominated
objects the self similarity can be related to the combined
influence of the standard violent relaxation and the regular
shape of the initial power spectrum of density perturbations.
However, within the observed objects it is often destroyed by
the impact of cooling of the baryonic component.

During previous discussion we used the redshift of object
formation, $z_f$, that is closely correlated with the central
DM density and its virial mass. It is important that $z_f$ is
clearly linked with the physical process of object
formation and in particular with the initial power spectrum. In
contrast, the popular use of the observed redshift of
clusters of galaxies $z_{obs}$ instead of the $z_f$ for estimates
of their virial characteristics is certainly an
incorrect procedure. The random character of the redshift
$z_{obs}\ll z_f$ is evident for galaxies. For CLASH clusters
discussed above this difference is also quite significant,
\[
\langle 1+z_f\rangle\approx 1.7\langle 1+z_{obs}\rangle\,.
\]
Introduction of $z_f$ instead of $z_{obs}$ requires more
detailed observations, but it  decreases
artificial variations of cluster characteristics and
in particular will change estimates of the redshift
evolution of the observed mass function of clusters.

The central pressure and temperature of relaxed objects are
determined mainly by the dynamical equilibrium of the
compressed, dominating DM component, but as was shown in section
3 the central entropy includes at least two components,
namely, the initial entropy of the compressed matter and
entropy generated in the course of violent relaxation. Our
results indicate that for clusters of galaxies the
contribution of the second component dominates. However, as
was discussed above and is seen from (\ref{smvir}) and
(\ref{s-infall}) for low mass early galaxies the contribution
of the initial perturbations can be more important.

\subsection{Properties of the baryonic component}

Properties of the baryonic component also strongly depend
upon the period of object formation. Thus, the first galaxies
such as dSph contain the baryonic component with very
low entropy determined by recombination of baryons (see,
e.g. Demia\'nski \& Doroshkevich 2014). This implies that for
these objects the central entropy of the baryonic component
is mainly generated in the course of their formation.

For $z_f\leq 10$ the entropy of slightly perturbed intergalactic
baryonic component can be partly related to the
progressive ionization and heating of the intergalactic gas by
the UV background. For redshifts $z_f\leq 3$ when strong
photo ionization of HeI and HeII is caused by the hard UV
radiation of quasars the temperature and entropy of such
baryons can be estimated as
\be
T_{b}\sim 0.7eV(1+z)^{6/7},\quad S_{b}\sim 18(1+z)^{-8/7}cm^2keV,
\quad z\leq 3\,.
\label{TSback}
\ee
This entropy strongly exceeds the entropy of the THING and
LSB galaxies and is similar to the observed entropy of less
massive clusters of galaxies. The Jeans mass of such baryonic
component increases up to
\[
M_J\sim 10^{10}M_\odot (1+z)^{0.2}\,.
\]
This shows that for less massive objects formed at $z_f\leq 3$
the baryonic fraction can be sharply suppressed.

\section{Shape of the power spectrum}

Observations of the relic microwave background radiation allow to
determine the shape of the initial power spectrum of density
perturbations at large scales starting from clusters of
galaxies (Komatsu et al. 2011; Ade et al. 2016). However,
information about the power spectrum at small scale and
composition and properties of dark matter is still missing.
In this section we consider in more details the function
$\Psi(M_{vir})$ introduced by (\ref{sz}) and link its mass
variations with the shape of the power spectrum.

\subsection{Theory versus observations}

Both simulations and observations show that characteristics
of the DM dominated halos are much more stable than
characteristics of the baryonic component, and they can be
used to characterize the small scale power spectrum of
density perturbations. Indeed our analysis performed in
sections 2,\,\,4\,\&\,5 shows that we can find a one  to  one
correspondence between the observed parameters of the DM
halos and the redshift of object formation,
$z_{f}$. According to the present day models of halo
formation (Press, Schechter, 1974; Peebles 1980; Bardeen
et al. 1986; Bond et al. 1991; Sheth\,\&\, Tormen 2002,
2004) the variations of redshift $z_f$ with the virial
mass of created objects characterize both the power
spectrum of density perturbations and the real period
of objects formation. As was discussed in section 2.3 in
these models the overdensity that divides the linear and
nonlinear periods of the object evolution is a weak function
of its virial mass. Therefore we can expect that
the function $\Psi(M_{vir})$ describing this division shows
also weak dependence upon the virial mass.

However, estimates (\ref{par1}),\,(\ref{grup1}),(\ref{par2}),
(\ref{par_kir}) and (\ref{udg}) demonstrate unexpectedly
significant mass dependence  of the function $\Psi(M_{vir})$\,
\[
 \langle\Psi(M)\rangle/\sigma_{8}\approx 0.4,\,0.74,\,1.1,
\,1.1,\,1.1 \,.
\]
This effect suggests a possible deviation of the small scale
power spectrum from that observed at larger scales by the
WMAP and Planck missions and now accepted in the $\Lambda$CDM
cosmological model.

\begin{figure}
\centering
\epsfxsize=7.cm
\epsfbox{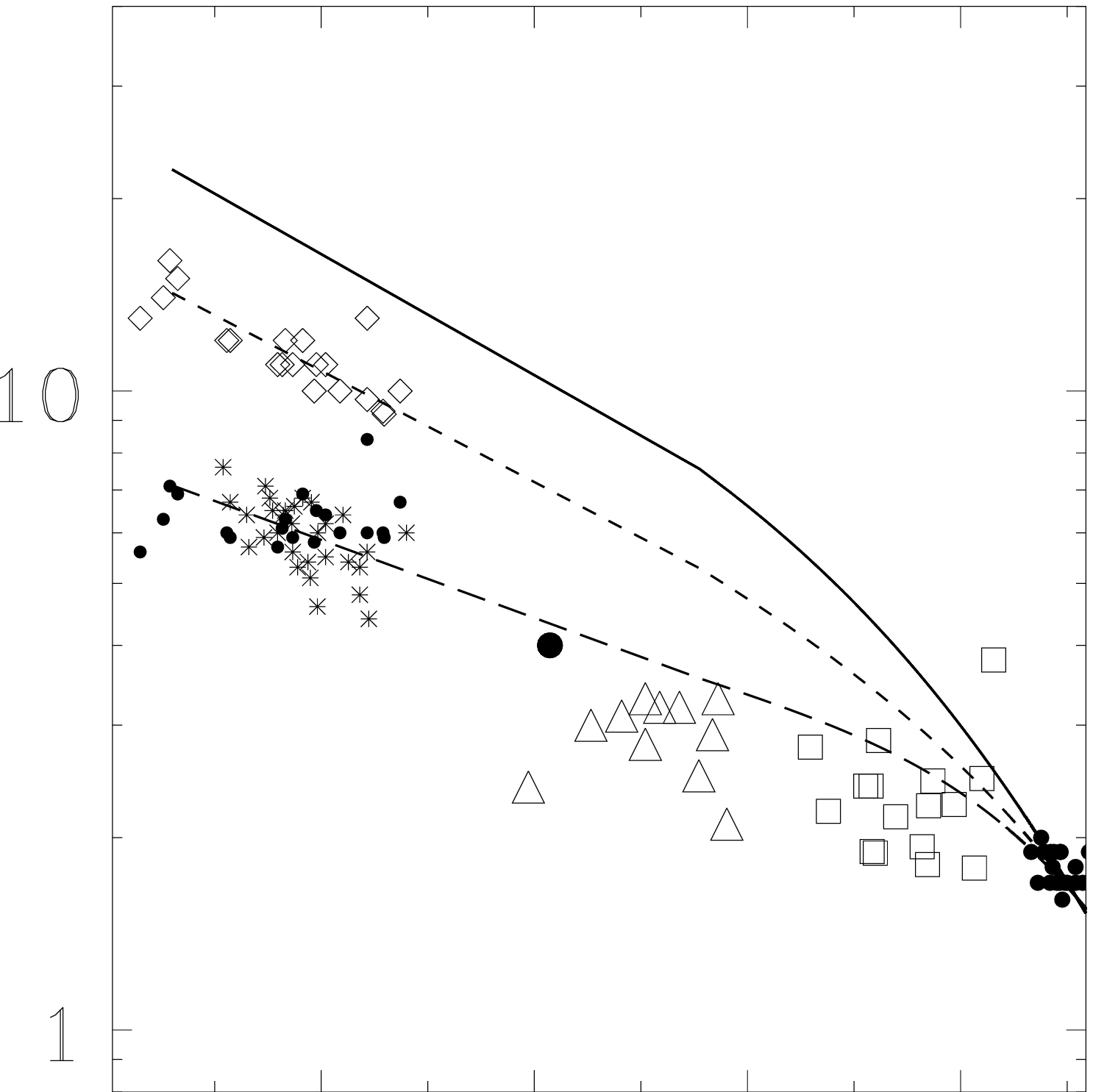}
\vspace{1cm}
\caption{Dispersion of the density perturbations $\sigma_m^*
=\sigma_m(M)/0.43$ (\ref{sigm},\,\ref{s**}) for the standard
$\Lambda$CDM power spectrum (\ref{sig_cdm}) and the combined
spectra (\ref{best1}) and (\ref{best2}) are plotted by solid,
long dashed and dashed lines vs. $M_{vir}/M_\odot$. Function
$B^{-1}(M_{vir})$ (\ref{bbz}) is plotted for the two samples
of dSph galaxies (left group of points and stars), for
Dragonfly 44 galaxy (filled circle) and for
CLASH clusters of galaxies (right group of points). For
THINGS and LSB galaxies and for groups of galaxies the
functions $B^{-1}(M_{vir})$ are plotted by triangles and squares.
For dSph galaxies the function $B^{-1}(M_{vir})$ obtained
assuming that  $\Phi(M)=1$ (\ref{phi_1}) is plotted by rhombus.
}
\label{sig6}
\end{figure}

This effect is illustrated in Fig. \ref{sig6} where the function
$B^{-1}(M_{vir})$ (\ref{bbz}) is plotted for the
samples of CLASH clusters, groups of galaxies and for the samples
of dSph, UDG, THING and LSB galaxies with both
choices of the function $\Phi(M_{vir})$ (\ref{phi_m}, \ref{phi_1}).
These functions are compared with the dispersion of
the density perturbations $\sigma_m(M_{vir})$ (\ref{sigm})
calculated for the standard $\Lambda$CDM power spectrum
(\ref{sig_cdm}). To more clearly represent the trend we plot
in Fig. \ref{sig6} the function
\be
\sigma_m^*=\sigma_m(M_{vir})/0.43\,, \label{s**}
\ee
that satisfies  the condition
\be
\Psi^*(M_{vir})=\sigma_m^*(M_{vir})B(z_f(M_{vir}))\approx 1\,,
\label{correc}
\ee
for the CLASH clusters of galaxies.
The strong differences between observations and expectations
of the standard CDM--like power spectrum are clearly seen
in Fig. \ref{sig6}.

At the same Figure the observed points $B^{-1}(M_{vir})$ are
well fitted by functions $\sigma_m^*=\sigma_m/0.43$
obtained for the more complex power spectra
\be
p_m(k)=0.1p_{cdm}(k)+0.9 p_{wdm}(k),\quad
\sigma_m^2=0.1\sigma_{cdm}^2+0.9\sigma_{wdm}^2\,,
\label{best1}
\ee
\be p_m(k)=0.4p_{cdm}(k)+0.6p_{wdm}(k),\quad
\sigma_m^2=0.4\sigma_{cdm}^2+0.6\sigma_{wdm}^2\,,
\label{best2}
\ee
Here $p_{cdm}(k)$ is the standard CDM power
spectrum. The function
\be
\sigma_{wdm}\approx 1.3\sigma_8/(1+0.05M_{12}^{0.4})\,,
\label{pwdm}
\ee
corresponds to the power spectrum describing the contribution
of the large scale perturbations only. The three functions
$\sigma_m^*(M_{vir})$ plotted in Fig. \ref{sig6} are identical
to each other for $M_{vir}\geq 10^{15}M_\odot$.

As an example of the power spectrum with damped small scale
part (\ref{pwdm}) we use the power spectrum of WDM particles
(Viel et al. 2005, 2013), namely,
\be
p_{wdm}(q)\approx p_{cdm}(q)[1+(\alpha_w q)^{2.25}]^{-4.46}\,,
\label{viel}
\ee
\[
q=\frac{k}{\Omega_mh^2},\quad
\alpha_w=6\cdot 10^{-3}\left(\frac{\Omega_mh^2}{0.12}
\right)^{1.4}\left(\frac{1 keV}{m_w}\right)^{1.1}\,,
\]
where $m_w\sim (50 - 100) eV$ and the comoving wave number
$k$ is measured in $Mpc^{-1}$. This value $m_w$ corresponds
to the damping scales $M_{dmp}$ and $D_{dmp}$
\be
\sigma_{wdm}(M_{dmp})=0.5\sigma_{wdm}(0),\quad
M_{dmp}=1.8\cdot 10^{15}M_\odot\,.
\label{md}
\ee
\[
D_{dmp}\sim (10 - 20) h^{-1}Mpc\,.
\]

Let us note however that in $p_{wdm}$ and $\sigma_{wdm}$ the
mass $m_w$ is only a  formal parameter allowing to introduce
the suitable damping scales $M_{dmp}$ and $D_{dmp}$ in the
function $\sigma_{wdm}$ (\ref{pwdm}) and such particles
need not really exist. The function $\sigma_{wdm}$
(\ref{pwdm}) is weakly sensitive to the shape of the
spectrum (\ref{viel}), when the suppression of power
occurs sufficiently rapidly. For example, the spectrum
with the Gaussian damping
\be
p_{wdm}(q)\approx p_{cdm}(q)[1+\exp(q^2/q_{dmp}^2)]^{-2}\,,
\label{gauss}
\ee
and a suitable value $q_{dmp}$ provides the same $\sigma_{wdm}$
(\ref{pwdm}). At the same time the CDM--like shape of the
small scale power spectrum is preserved in (\ref{best1}),
(\ref{best2}), which is consistent with similarity of the
expressions (\ref{sig_cdm}) and (\ref{zm}) for $M_{12}\leq 1$.
It is apparent that such decrease of the amplitude of small
scale perturbations eliminates discrepancy between estimates
(\ref{par3},\,\ref{par1},\,\ref{grup1}\,\&\,\ref{par2})
of the function $\Psi(M_{vir})$.

The difference between the power spectra (\ref{best1}) and
(\ref{best2}) is caused by the choice of the function
$\Phi(M_{vir})$. But as is seen from Fig. \ref{sig6} and from
comparison of (\ref{par1}) and (\ref{par4}) even for the
case (\ref{best2}) the differences between the observed
values $B^{-1}(M_{vir})$ and $\sigma_{cdm}$ for the $\Lambda$CDM
power spectrum remain quite large. These results show that in
contrast with some theoretical expectations (see, e.g. Ellis
et al. 2016) the power spectrum at small scales should be
significantly suppressed as compared with the standard
$\Lambda$CDM model, but without changing its shape.

Of course these results cannot be considered as a strong
indication that the standard $\Lambda$CDM model should be
modified, but they demonstrate again some intrinsic problems of
this model. The strong links of DM characteristics of observed
relaxed objects with the initial power spectrum and cosmological
model is well known. Results presented in Figs. \ref{fprs} and
\ref{sig6} demonstrate the regular correlations of properties
of DM component with the virial masses of the relaxed objects.
These correlations can be successfully interpreted in the
framework of the standard model of halo formation based on the
excursion set approach, but applied for the more complex power
spectrum. However, these unexpected results should be tested with
suitable set of high resolution simulations before their real
status will be accepted.

Of course the observational base used in our discussion is very
limited and it should be extended by more observations
of objects with masses $M\leq 10^{12} M_\odot$, what may be crucial
for determination of the shape of the initial power
spectrum, the real composition of dark matter and even the models
of inflation. Unfortunately more or less reliable
estimates of the redshift $z_{f}$ can be obtained mainly for the
DM dominated objects or for objects with clearly
discriminated impact of DM and baryonic components. Here we use
results obtained for 11 THINGS and LSB galaxies (de
Blok et al. 2008; Kuzio de Naray et al. 2008), but even for them
both the virial masses and redshift $z_f$ are strongly
underestimated. The most promising results can be obtained for
the mentioned in section 5.7 population of Ultra Diffuse
Galaxies  now represented by the galaxy Dragonfly 44 only
(see Figs 1\,\&\, 2). We hope that the list of possible
appropriate candidates will be extended.

Next important problem is the reliability  of the approach
used in our analysis and obtained results. It depends upon
the representativity of the observational data and is
moderate because of the very limited available data and
their significant scatter. Progress achieved during last
years makes it possible to begin discussion of this problem,
but the available data allow only qualitative character
of such discussion. Indeed, the problem of estimates
of the mass, density and other parameters of the observed
objects is quite complex, methods used for such estimates
are very rough and model dependent and their reliability
is limited. Moreover the influence of the baryonic component
increases scatter of the measured parameters and makes it
difficult to estimate the real precision of our results.
However, we hope that because of the high importance of
this problem such observations will be extended and their
precision improved.

\subsection{'Too Big To Fail' approach}

Essential support for our inferences comes from comparison
of the circular velocities of simulated low mass galaxies
and the observed dSph satellites of Milky Way and Andromeda
(Boylan--Kolchin et al. 2012; Garrison--Kimmel et al.,
2014a,b; Tollerud et al. 2014; Hellwing et al. 2015; Brook,
\& Cintio 2015). It demonstrates that the
circular velocities of objects in simulations performed
with the standard $\Lambda$CDM power spectrum reproduce
observations for more massive objects, but regularly
overestimate the observed circular velocities for less
massive objects.

It seems that this discrepancies can be related to
efficiency of the environmental processes and complex
evolutionary history in the context of the hierarchical
formation of objects. This  problem is discussed by Wetzel
et al. 2015. But the similar effects are observed also
for galaxies in the Local Group, where, in particular,
there is the deficit of low mass galaxies in observations
as compared with predictions of the standard simulations
(Klypin et al. 2015). These discrepancies are smaller
for simulations performed with suppressed power at small
scale (Garrison--Kimmel et al., 2014b; Marsh \& Silk 2014).
However, as was noted above simulations with the standard
WDM power spectrum cannot reproduce observations.

These results show that more complex improvements
of the standard theoretical models are required. But
these indications of qualitative disagreements between
standard simulations and observations are not followed
by quantitative estimates of required corrections.

\subsection{Extended $\Lambda$CDM model.}

The important result of our analysis is the indication of some
weakness of the standard $\Lambda$CDM cosmological model and
great potential of models with more complex shape of the power
spectrum. However, our approach indicates that the basic
element of such modified power spectrum could be the large
contribution ($\sim 70\%$) of the spectrum often associated
with particles with large damping scale (\ref{md}). Nonetheless
such partial damping of the power spectrum does not
forbid formation of less massive objects, but rather decreases
their rate of formation relatively to the standard
$\Lambda$CDM -- like power spectrum. It is noteworthy that after
30 years of absolute domination of the CDM model we
begin to consider more complex versions of the HDM models
(Doroshkevich et al. 1980a,b).

It is important that the observed characteristics of objects
can be described by the linear combination of power
spectra and so by the damping scales that in turn depend both
upon particle masses and velocities. This means that the
construction of an adequate complex cosmological model should
start with complex inflation models or/and include
discussion of DM composition with estimates of the actual
damping scales and transfer functions allowing also for
linear evolution of perturbations at $z\leq z_{eq}$.

For the models with one component DM the assumption of the more
complex primeval power spectrum is required, which
implies more complex inflation models (or some other equivalent
assumption). However, in this case the condition
$\Psi(M)\approx const.$ (\ref{const}) is carried out at
$z\leq z_{eq}$ and expressions (\ref{best1}) -- (\ref{md})
correctly describe the subsequent process of objects formation.
In contrast, for the more interesting models with multi
component composition of DM this description should be corrected.
As is shown in Appendix B in such models the complex
evolution of perturbations leads to deformation of the
transfer function even at small redshifts and to redshift
dependence of the function $\Psi(M,z)$ (\ref{sz}). In turn this
complicates the description of the sequential object
formation. For such models our estimates (\ref{best1}) --
(\ref{md}) become qualitative and approximate. They should be
improved by comparison with suitable simulations.

It seems that the simplest such model is a suitable
combination of very massive and light particles,
what leads to a complex shape of the transfer function
after the period of recombination (see, e.g. discussion
in Boyarsky et al. 2009a,b). However, abilities of this
simple model are strongly restricted by
the period of cosmic nucleosynthesis. In the standard
model with the three oscillating neutrinos the effective
number of light particles is
\[
N_{eff}=3.046\,,
\]
(Mangano et al. 2006) and the small difference between this
value and latest estimates (Ade et al. 2016) makes
theoretical models more complex. Thus, for discussions of the
sterile neutrinos with mass $m\sim 1 eV$ it is necessary
to assume that they cannot be in thermal equilibrium before
the period of cosmic nucleosynthesis (see, e.g., Abazajian
2012, 2015; Kopp et al., 2013; Zysina, Fomichev \& Khruschov
2014). Recently  properties of sterile
neutrinos were strongly restricted by the Ice Cube observations (Aartsen
et al. 2016).

Of course, all these inferences are very preliminary. Thus,
here we use the power spectra with the transfer function
(\ref{viel})\, or with Gaussian damping (\ref{gauss}).
More refined description of the power spectrum and
specially further progress in the observations of DM
dominated objects with suitable virial masses will improve
the best model parameters and estimates of the masses and
composition of the DM particles. Nonetheless even today
replacement of some fraction of CDM particles by particles
with  $m_w\sim 3 - 10\, keV$ can be considered. Modification
of the spectrum by such particles, identified now (according
to majority preference) as the sterile neutrinos, can be
included in (\ref{best1}) as a third component without
noticeable changes of Fig. \ref{sig6}. Indeed, the available
sample of the observed DM dominated objects does not yet
allow to make any far--reaching conclusions about the
actual properties of DM particles.

The widely discussed controversial characteristics of DM
dominated objects such as the core--cusp problem, or number of low
mass satellites (see, e.g., Bovill \& Ricotti, 2009;
Koposov et al., 2009; Walker, \& Penarrubia, 2011;
Boylan-Kolchin et al., 2012; Penarrubia et al. 2012;
Governato et al. 2012; Sawala, 2013; Teyssier et al.
2013; Laporte et al. 2013; Collins et al. 2014; Miller
et al. 2014) depend upon the dissipative scale and the
shape of the power spectrum. This fact supports the hope
that these problems also can be successfully resolved in
the framework of  more complex cosmological models (see,
e.g., Pilipenko et al. 2012).

Accurate simulations with various power spectra and/or complex DM
composition are required before we will have next generation
of cosmological models with more reliable properties of the
DM component. These expectations are supported by moderate
results of the first published simulations of the clearly
artificial WDM cosmological models (Maccio 2012, 2013;
Angulo, Hahn, Abel, 2013; Schneider, Smith \& Reed, 2013;
Wang et al. 2013; Libeskind et al. 2013; Marcovi$\breve{c}$
\& Viel 2013; Schultz et al. 2014; Schneider et al. 2014;
Dutton et al. 2014).

\section{Conclusions}

In this paper we  propose a new approach that could solve
some important problems of modern cosmology:
\begin{enumerate}
\item{} Explain the self similarity of the internal structure
of virialized DM dominated objects in a wide range of masses, which
is manifested as the regular dependence of the
central pressure, density, entropy and the epoch of halos formation
(\ref{psm}, \ref{ffit}) upon the virial mass of objects.
\item{} Unexpectedly weak dependence of the DM pressure and surface
density of cores of DM dominated virialized objects on their virial
masses.
\item{}Establish the real composition of dark matter
and the shape of the small scale power spectrum of density
perturbations\,.
\end{enumerate}

Summing up let us note that the proposed approach allows us to
consider and to compare properties of the observed DM
dominated objects in an unprecedentedly wide range of masses
$10^6\leq M_{vir}/M_\odot\leq 10^{15}$. The main results
presented  in Table 1  and plotted in Fig. \ref{fprs}
unexpectedly favor the regular self similar character of the
internal structure of these objects, what confirms the main
expectations of the NFW model and the regular character of
the power spectrum without sharp peaks and deep troughs.

On the other hand as is seen from Fig. \ref{sig6} our
results favor the models with more complex power spectrum
with significant excess of power at cluster scale and/or,
alternatively, deficit of power at dwarf galactic scales.
Both alternatives seem to be quite important and call for
more detailed observational study of DM dominated objects.
Our results supplement the traditional investigations of
galaxies at high redshifts (McLeod et al. 2015; Robertson
et al. 2015; Bouwens et al., 2015a,b; Ellis et al. 2016).

Here we consider a phenomenological model of the power spectrum
that is composed of fraction $g_{cdm}\sim 0.1 - 0.4$ of
the standard CDM spectrum and fraction $g_{wdm}\sim 0.9 - 0.6$ of
the HDM spectrum with the transfer function
(\ref{viel}). Unexpectedly in the spectrum (\ref{best1}) the
contribution of low mass particles with relatively large
damping scale (\ref{md}) dominates, what can be considered as
reincarnation in a new version of the earlier rejected
HDM model. Further progress can be achieved with more complex
models of the power spectrum and/or more realistic
transfer functions instead of (\ref{viel},\, \ref{gauss}), what
implies also more realistic complex composition of the
DM component. In particular
it is possible to discuss an excess of power
localized at small scale. In some respect such an excess of power
reminds the isocurvature models (see, e.g. Savelainen
et al. 2013) with similar predictions and problems. However, all
such problematic multi parametric proposals should be
considered in the context of the general cosmological and
inflationary models.

The approach used in this paper for discussion of composition
and properties of the DM particles is very indirect. We
consider effects of strongly nonlinear multistep evolution of
perturbations resulting in observed properties of the
relaxed objects. The nonlinear evolution usually leads to a
strong loss of information about the primeval perturbations
and composition of the DM component. These losses are further
enhanced by the masking effects of dissipative evolution
of the baryonic matter. To reveal the missing impact of the
DM composition and primeval perturbations we discuss the
most conservative characteristics of the observed DM dominated
objects by comparing these data with numerical
simulations majority of which are now focused on studying
evolution of the standard $\Lambda$CDM model.

We believe that more detailed conclusions will be made
on the basis of special simulations and after accumulation
of more representative observational sample of high precision
measurements of properties of DM dominated objects.

\subsection{Acknowledgments}
This paper was supported in part by  the grant of the
President of RF for support of scientific schools
NSh-6595.2016.2. We thank S. Pilipenko, A. Klypin,
B. Komberg, A. Saburova, M. Sharina and R. Ruffini for
useful comments. Our analysis would not be possible
without detailed measurements presented in cited papers.
We thank the referee for constructive comments that
significantly improved the text and eliminated many
unclear statements.

\begin{appendix}

\section{Entropy of the relaxed DM halos}

Here we estimate the two main components of entropy of DM relaxed
halos. First is the entropy generated during compression
and violent relaxation of a spherical DM cloud at rest (Peebles
1967; 1980). The other one includes entropy generated by random
motions associated with initial density perturbations.

\subsection{The mean entropy of halos: Peebles model}

In this model the evolution and relaxation of DM halos with
zero initial entropy is considered. Thus, this model
demonstrates that some entropy is really generated within
collapsed and relaxed DM halos even without any initial
entropy. To obtain rough estimate of this entropy we
use the well known correlations between the gravitational
$U$ and internal $W$ energy of virialized object,
\[
\theta_{vir}=\frac{U}{W}\approx \frac{GM_{vir}}{R_{vir}
T_{x}}=const\,,
\]
where $T_x$ is the halo temperature and the value of
the $const$ depends on the internal structure of the object.
Thus, for 180 observed clusters of galaxies with masses $10^{13} \leq
M_{vir}/M_\odot\leq10^{15}$ we have
\be
\theta_{vir}=\frac{M_{vir}}{10^{12}M_\odot}\frac{1 Mpc}{R_{vir}}
\frac{1 kev}{T_x}\approx 72(1\pm 0.08)\,.
\label{MRT}
\ee
Using this relation and the estimates of the mean
density of relaxed DM halos (\ref{pmod}) we get
\[
R_{vir}\approx \frac{0.33MpcM_{12}^{1/3}}{(1+z_f)\Phi^{1/3}},
\quad T_x\approx 0.04M_{12}^{2/3}(1+z_f)
\frac{72\Phi^{1/3}}{\theta_{vir}} keV\,,
\]
\be
\langle S_{vir}\rangle\approx \frac{T_x}{\langle
n_{DM}\rangle^{2/3}}\approx 10^{-2}\frac{M_{12}^{2/3}
\mu_{DM}^{2/3}}{(1+z_f)\Phi^{1/3}}\frac{72}{\theta_{vir}}cm^2keV\,,
\label{Svir}
\ee
\[
\mu_{DM}=m_{DM}/m_b\,.
\]

Here $n_{DM}$ is the mean number density of DM particles
in the halo, $m_{DM}$ and $m_b$ are the masses of DM
particles and baryons. Using the relation (\ref{zm}) we get
\be
\langle S_{vir}(M)\rangle\approx
4.3M_{12}^{0.74}cm^2keV \frac{\mu_{DM}^{2/3}}{\Phi^{1/3}}
\frac{\eta_f}{4.1} \frac{72}{\theta_{vir}}\,,
\label{asmvir}
\ee
which is comparable with observational estimates (\ref{ffit}).
However, these estimates apply mainly to peripheries of
halos with large entropy.

\subsection{The entropy of halos: contribution of
random motions}

According to the Zel'dovich approximation and in accordance
with simulations based on the $\Lambda$CDM cosmological
model the velocity dispersion along any principle direction is
(see, e.g., Demia\'nski et al. 2011)
\be
\sigma_U=\frac{H(z)}{1+z}\beta(z)B(z)\frac{\sigma_s}{ \sqrt{3}}
\approx\frac{350km/s}{\sqrt{1+z}}
\frac{\sigma_8}{0.8}\Theta_m^{-1}\,,
\label{velo-1}
\ee
where the functions $H(z)\,\&\,B(z)$ were introduced in
(\ref{basic}),
\[
\sigma_s^2=\frac{1}{2\pi}\int_0^\infty p(k)dk,\quad
\beta(z)=-\frac{1+z}{B(z)}\frac{dB(z)}{dz}\,,
\]
\[
\sigma_s\approx 9.3h^{-1}Mpc\Theta_m^{-1}\sigma_8/0.8\,.
\]

However, all actual velocities within the
compressed clouds do not exceed 50 - 100 km/s because the
velocity (\ref{velo-1}) is dominated by the random
velocity of the halo as a whole. To obtain a rough estimate
of random motions of particles accumulated within the
compressed halo we can use the relation
\be
\sigma_v^2(M_{vir},M_{c})=\frac{1}{2\pi\sigma_s^2}\int_0^\infty
dk\,p(k)\Xi(k,M_{vir},M_c)\,,
\label{sig_vv}
\ee
\[
\Xi(k,M_{vir},M_c)=1+W^2(k,M_{vir})-2W(k,M_{vir})W(k,M_c)\,,
\]
where the function $W(k,M)$ was introduced in (\ref{sigm}),
$M_{vir}$ is the virial mass of a halo and $M_c\leq M_{vir}$
is the mass of the central core of this halo. The function
$\sigma_v^2(M_{vir},M_c)$ rapidly increases with the virial mass
$M_{vir}$, but only weakly depends upon the mass $M_c$. For
$M_c=0$ it is well fitted by the expression
\be
\sigma_v^2(x)=1.5\cdot 10^{-2}x^{0.6}/(1+0.2x^{0.35})\,.
\label{fit-vgm}
\ee

Using these results we can roughly estimate the temperature
$T_{rnd}$ and the entropy $S_{rnd}(M_{vir})$ of the
compressed matter within a halo with a mass $M_{vir}$ formed at
the redshift $z_f$ as
\[
T_{r}(M_{vir})\approx \frac{3m_{DM}\sigma_U^2}{2}g_{r}\sigma_v^2(M_{vir})
\approx \frac{30M_{12}^{0.6}\mu_{DM}}{1+z_f}g_{r}eV\,,
\]
\be
S_{r}=\frac{T_{r} m_{DM}^{2/3}}{\langle\rho_m\rangle^{2/3}}
g_{r}\approx 220cm^2keV\frac{M_{12}^{0.6}g_{r}}{(1+z_f)^3}
\mu_{DM}^{5/3}\,,
\label{s-infall}
\ee
where $g_{r}$ is a random factor that determines the fraction
of energy of random motions (\ref{sig_vv}) accumulated by
the halo. Using the correlation between the redshift $z_f$
and the virial mass of halo (\ref{zm}) we get
\be
S_{r}(M_{vir})\approx 6.4M_{12}^{0.81}\mu^{5/3}g_{r}(\eta_f/4.1)^3
cm^2keV\,,
\label{as2i}
\ee
which is comparable with observational estimates (\ref{ffit}).
It seems that this channel of entropy generation is
more important for less massive halos.

\subsubsection{ The entropy of intergalactic gas}

After reionization the intergalactic  gas is heated
by the UV radiation and its temperature $T_b$ is close to
$10^4K$ with large spatial variations determined by the
inhomogeneities of the UV foreground. The entropy of this gas
can be estimated as
\be
\langle S_{bar}\rangle\approx\frac{\langle T_b\rangle}{\langle
n_b\rangle^{2/3}} \approx\frac{4 cm^2keV}{(1+z_f)^2}\sim
0.16M_{12}^{0.15}cm^2keV \,.
\label{sbar}
\ee

\section{Evolution of perturbations in two component
non relativistic medium}

The evolution of perturbations in two component non relativistic
medium were considered by Grishchuk \& Zel'dovich
1981. Let us assume that the first component is composed of
light nonrelativistic particles with density
$\rho_1=\alpha_1\langle\rho(z)\rangle\propto (1+z)^{-3}$, the
pressure $p_1\propto\rho_1^{5/3}$ and the sound speed
$\beta_1\propto 1+z$. The second component is composed of
massive particles with density $\rho_2
=\alpha_2\langle\rho(z)\rangle\propto (1+z)^{-3}$,
$\alpha_2=1- \alpha_1$ and negligible pressure and sound speed,
$p_2=0,\,\, \beta_2=0$. In this case the evolution of the
density perturbations is described by the equations
\be
y^2\delta_1''+\frac{y}{2}\delta_1'-\frac{3}{2}[\alpha_1
\delta_1 +\alpha_2\delta_2-\kappa^2y^2\delta_1]=0\,,
\label{eq1}
\ee
\[y^2\delta_2''+\frac{y}{2}\delta_2'-\frac{3}{2}[\alpha_1\delta_1
+\alpha_2\delta_2]=0\,,
\]
where $'$ denotes derivative with respect to the redshift
$y=1+z$, $\delta_1=\delta\rho_1/\rho_1,\,\delta_2=\delta
\rho_2/\rho_2,$ are density perturbations in components 1 \& 2,
$\kappa^2=\beta_1^2k^2$ and $k$ is the comoving wave number.

For large scales ($\kappa\rightarrow 0$) we get the usual result for
the increasing ($c_1$) and decreasing ($c_2$) adiabatic modes
\be
\alpha_1\delta_1+\alpha_2\delta_2\approx c_1(1+z)^{-1}+c_2(1+z)^{3/2}\,,
\label{amode}
\ee
and for two entropic modes we get
\be
\delta_1-\delta_2\approx c_3+c_4(1+z)^{-1/2}\,.
\label{emode}
\ee
On the other hand for the small scale perturbations with
$\alpha_1\delta_1\ll\alpha_2\delta_2\ll\kappa^2y^2\delta_1$ we get
\be
\delta_2\approx c_5(1+z)^{-\gamma},\quad \gamma=(\sqrt{1+
24\alpha_2}-1)/4\,,
\label{ad2}
\ee
\[
\delta_1\approx c_6(1+z)^{1/4}J_{1/4}[\kappa (1+z)]\propto (1+z)^{-1/4}
\]
where $J_\nu$ is the Bessel function. For $\alpha_2=1$ we get from
(\ref{ad2}) again that $\delta_2\propto (1+z)^{-1}$, but
$\gamma\leq 1$ for $\alpha_2\leq 1$.

These results demonstrate that for the two component medium
with strongly different masses of particles the spectrum of
perturbations is distorted. Thus, if for large scale
perturbations the amplitude grows as $\delta_{1,2}\propto
(1+z)^{-1}$, then for small scale perturbations we get, even
for heavy particles, that $\delta_2\propto (1+z)^{-\gamma}$
(\ref{ad2}) with $\gamma\leq 1$. For the critical value $\alpha_2
=1/8$ we get $\gamma=1/4$, that is the same as the exponent
for the small scale perturbations of the other component described
by $\delta_1$. Thus, in this case perturbations in both media evolve in
the same way.

\end{appendix}

\bibliographystyle{aa}

\begin{thebibliography}{}

\expandafter\ifx\csname natexlab\endcsname\relax\def\natexlab#1{#1}\fi

\bibitem[]{Aartsen}
Aartsen et al. 2016, Phys. Rev.Lett. 117, 071801

\bibitem[]{Aba 2012}
Abazajian, K., Acero, M., Agarwalla, S., et al., 2012, arXiv:1204.5379

\bibitem[]{Aba 2015}
Abazajian, K., Arnold, K., Austerman, J., et al, 2015, Aph, 63, 66

\bibitem[{Abbott, T., Abdalla, F., Allam, S., et al. 2015}]{Abb 2015}
Abbott, T., Abdalla, F., Allam, S., et al. 2015, arXiv:1507.05552

\bibitem[]{Abd 2015}
Abdallah, J., Araujo, H., Arbey, A., et al., 2015, PDU, 9, 8

\bibitem[]{adam}
Adam, R., Comis, B., Macias-Perez, J., et al., 2015, A\&A, 576, 12

\bibitem[]{Ade 2016}
Ade, P., Aghanim, N., Arnaud, M., et al. 2016, A\&A, 594, 13
\bibitem[]{Agnese 2014}
Agnese, R., Anderson, A., Asai, M., et al. 2014, PhRvL, 112x1302

\bibitem[]{Anglo}
Angloher, G., et al., 2013, Eur.Phys.J., C72, 1971

\bibitem[]{Ander}
Anderhalden, D., Schneider, A., Maccio, A., Diemand, J.,
Bdertone, G., 2013, LCAP, 03, 014

\bibitem[]{angulo 2013}
Angulo, R., Hahn, O., Abel, T., 2013, MNRAS, 434, 3337

\bibitem[]{Arnaud 2005}
Arnaud, M., Pointecouteau, E., Pratt, G., 2005, A\&A, 441, 893

\bibitem[]{Arnaud 2010}
Arnaud, M., Pratt, G., Piffaretti, R. et al.,
2010, A\&A, 517, 92;

\bibitem[]{balas}
Balazs, L., Bagoly, Z., Hakkila, J. et al.
2015, MNRAS, 452, 2236

\bibitem[]{BBKS}
Bardeen~J.M., Bond~J.R., Kaiser~N., \& Szalay~A., 1986,
ApJ, 304, 15

\bibitem[]{Bar}
Baring, M., Ghosh, T., Queiroz, F., Sinha, K., 2016, PhRvD, 93.103009;

\bibitem[]{batta}
Battaglia, N., Bond, R., Pfrommer, C., Sievers, J., 2015, ApJ, 806, 43

\bibitem[]{battye-14}
Battye, R., Moss, A., 2014, PhRvL., 112e1303

\bibitem[]{battye}
Battye, R., Charnock, T., Moss, A., 2015, PhRvD..91j3508

\bibitem[]{Beasley}
Bennet C., et al., 2003, ApJS,, 148, 1

\bibitem[]{Benn}
Beasley, M., Romanowsky, A., Pota, V., et al., 2016, ApJ, 819, L20

\bibitem[]{Bere-15}
Berezhiani, Z., Dolgov, A., Tkachev, I., 2016, PhRvD, 92f1303;

\bibitem[]{Bere-15}
Berezinsky, V., Dokuchaev V., Eroshenko Yu., 2014,  Phys. Usp. 57, 1

\bibitem[]{Bern}
Bernabei, R., Belli, P., Capella, F., et al., 2010, Eur.Phys.J., C67, 39

\bibitem[]{Bhatta-13}
Bhattacharya, S., Habib, S., Heitmann, K., Vikhlinin, A., 2013,
ApJ, 766, 32

\bibitem[]{Blen}
Blennow, M., Ccoloma, P., Fernandez--Martinez, E., Machado, P.,
Zaldivar, B., 2016, JCAP, 04, 015

\bibitem[]{Blin}
Blinnikov, S., 2014, Phys. Usp. 57 183

\bibitem[]{bond 1991}
Bond, R., Cole, S., Efstathiou, G., Kaiser, N., 1991, ApJ,
379, 440

\bibitem[]{borsanyi 2016}
Borsanyi, S., Dierigl, M., Fodor, Z., et al., 2016, PhLB, 752, 175

\bibitem[]{bovill 2009}
Bovill, M, Ricotti, M., 2009, ApJ, 693, 1859

\bibitem[]{bouwens 2015}
Bouwens, R., Illingworth, G., Oesch, P. et al.,,
2015a, ApJ, 811, 140

\bibitem[]{bowens 2015}
Bouwens, R., Oesch, P., Labbe, I., et al., 2015b,
arXiv:1506.01035

\bibitem[]{boy 2009}
Boyarsky, A., Lesgourgues, J., Ruchayskiy, O., Viel, M., 2009a,
JCAP, 05, 012

\bibitem[]{bo 2009}
Boyarsky, A., Lesgourgues, J., Ruchayskiy, O., Viel, M., 2009b,
 PhRvL, 102t1304

\bibitem[]{boya 2009}
Boyarsky, A., Ruchayskiy, O., Shaposhnikov, M., 2009c,
ARNPS, 59, 191

\bibitem[]{boyl 2012}
Boylan-Kolchin, M., Bullok, J., Kaplinghat, M., 2012, MNRAS,
422, 1203

\bibitem[]{Branchesi}
Branchesi, M., Gioia, I., Fanti, C.,Fanti, R., 2007, A\&A,
472, 739

\bibitem[]{Bredd}
Breddels M., \& Helmi, A., 2014, ApJ, 7911, 3

\bibitem[]{Brooks-st}
Brook, C., Di Cintio, A., Knebe, A., Gottlöber, S.; Hoffman, Y.,
Yepes, G., Garrison-Kimmel, S., 2014, ApJ., 784, L14

\bibitem[]{Brook}
Brook, C., Cintio, A., 2015, MNRAS, 450, 3920

\bibitem[]{Brooks}
Brooks, A., \& Zolotov, A., 2014, ApJ, 786, 87

\bibitem[]{Bryan-n}
Bryan, G., Norman, M., 1998, ApJ, 495, 80

\bibitem[]{Bulbul}
Bulbul, E., Markevitch, M., Foster, A., Smith, R., Loewenstein,
M., Randall, S., 2014, ApJ, 789, 13

\bibitem[]{Bull}
Bullock, J., Kolatt, T., Sigad, Y. et al.,
 2001, MNRAS, 321, 559

\bibitem[]{Bur}
Burenin, R., Vikhlinin, A., 2012, Astronomy Lett., 38, 1

\bibitem[]{Burk}
Burkert~A., 1995, ApJ, 447, L25

\bibitem[]{Burk15}
Burkert~A., 2015, ApJ, 80, 158

\bibitem[]{Chan 2015}
Chan, T., Kereš, D., Oñorbe, J., et al., 2015, MNRAS, 454, 2981

\bibitem[]{Chemin 2011}
Chemin, L., de Blok, W., Mamon, G., 2011, AJ, 142, 109

\bibitem[]{Coll 2014}
Collins, M., Chapman, S., Rich, R., et al., 2014, ApJ, 783, 7

\bibitem[]{croston}
Croston, J., Pratt, G., Bohringer, H., et al., 2008, A\&A, 487, 431

\bibitem[]{deblok}
de Blok, W., Walter, F., Brinks, E., Trachternach, C., Oh,
S.-H., Kennikutt, R., 2008, AJ, 136, 2648

\bibitem[]{demianski 1999}
Demia\'nski, M., Doroshkevich, A., 1999, ApJ, 512, 527

\bibitem[]{demianski 2004}
Demia\'nski, M., Doroshkevich, A., 2004, A\&A, 422, 423

\bibitem[]{dor 2007}
Demia\'nski, M., Doroshkevich A., 2007, PhRvD.,7513517

\bibitem[]{demianski 2011}
Demia\'nski, M., Doroshkevich, A., Pilipenko, S., Gottl\"ober, S.,
2011, MNRAS, 414, 1813.

\bibitem[]{dor 2014}
Demia\'nski, M., Doroshkevich A., 2014, MNRAS, 439, 179.

\bibitem[]{di 2007}
Diemand, J., Kuhlen, M., Madau, P., 2007, ApJ, 667, 859

\bibitem[]{diem 2013}
Diemer, B., More, S., Kravtsov, A., 2013, ApJ, 766, 25

\bibitem[]{diem 2014}
Diemer, B., Kravtsov, A., 2014, ApJ, 789, 1 

\bibitem[]{dkk 1980}
Doroshkevich, A., Khlopov, M., Sunyaev, R., Zel'dovich, Ya.,
1980a, SvAL,  6,  252

\bibitem[]{dkk 1981}
Doroshkevich, A., Khlopov, M., Sunyaev, R., Szalay, A.,
Zel'dovich, Ya., 1980b, Proc. 10th Texas Symposium on Relativistic
Astrophysics, Ann.New York Acad. Sci., 375, 32.

\bibitem[]{dor 1981}
Doroshkevich A., Zel'dovich Ya., 1981, JETP, 80, 801
\bibitem[]{dk 1984}
Doroshkevich, A., Khlopov, M., 1984, MNRAS, 211, 277.

\bibitem[]{dkk 1986}
Doroshkevich, A., Khlopov, M., Kotok, E., 1986, SvA., 30, 251

\bibitem[]{dkk 1988}
Doroshkevich, A., Klypin, A., Khlopov, M., 1988, SvA., 32, 127

\bibitem[]{dor 2011}
Doroshkevich A., Verkhodanov, O., 2011, PhRvD, 83d3002

\bibitem[]{dor 2012}
 Doroshkevich A., Lukash, V., Mikheeva, E., 2012, PhyU., 55, 3

\bibitem[]{Dutton 2014}
Dutton, A., Maccio, A., 2014, MNRAS, 441, 3359

\bibitem[]{einasto}
Einasto, M., 2014, arXiv:1409.1347

\bibitem[]{eisen}
Eisenstein, D., Hu, W., 1998, ApJ., 496, 605

\bibitem[]{ellis}
Ellis, J., Garcia, M., Nanopulos, D., Olive, K., 2016, CQGra, 33i4001;

\bibitem[]{enqv 2015}
Enqvist, K., Nadathur, S., Sekiguchi, T., Takahashi, T., JCAP, 09, 067

\bibitem[]{essig 2013}
Essig, R., Jaros, J., Wester, W., et al., 2013, arXiv:1311.0029

\bibitem[]{feng 2010}
Feng, J., 2010, ARA\&A, 48, 495

\bibitem[]{hunter 2013}
Ferrer, F., Hunter, D., 2013, JCAP, 09, 005

\bibitem[]{Fillmore-Goldreich}
Fillmore~J.A., \& Goldreich~P., 1984, ApJ, 281, 1

\bibitem[]{Fitz}
Fitts, A., et al. arXiv:1611.02281

\bibitem[]{gar14a}
Garrison-Kimmel, S., Boylan-Kolchin, M., Bullock, J., Kirby, E.,
2014a, MNRAS, 444, 222

\bibitem[]{gar14b}
Garrison-Kimmel, S., Horiuchi, S., Abazajian, K., Bullock, J.,
Kaplinghat, M., 2014b, MNRAS, 444, 961

\bibitem[]{gar14c}
Garrison-Kimmel, S., Boylan-Kolchin, M., Bullock, J., Lee, K.,
 2014c, MNRAS, 438, 2578

\bibitem[]{Gor}
Gorbunov, D., 2014, Phys. Usp. 57 503

\bibitem[]{Gover}
Governato, F., Zolotov, A., Pontzen, A., et al., 2012,
MNRAS, 422, 1231

\bibitem[]{grishchuk}
Grishchuk, L., \& Zel'dovich, Ya.,
1981, Astron. Zh., 56, 472

\bibitem[]{Gurevich}
Gurevich, A., Zybin, K., 1995, Phys.Usp., 38, 687

\bibitem[]{hell 2015}
Hellwing, W., Frenk, C., Cautun, M., et al. 2016, MNRAS, 457, 3492

\bibitem[]{hin 2013}
Hinshaw G., Larson, D., Weiland, J., et al., 2013, ApJS, 208, 19

\bibitem[]{hor 2015}
Horiuchi, S., Bozek, B., Abazajian, K., Boylan-Kolchin, M.,
Bullock, J., Garrison-Kimmel, S., Onorbe, J., 2016, MNRAS, 456, 4346

\bibitem[]{hor 2015}
Horvath, I., Bagoly, Z., Hakkila, J., Toh, L., A\&A, 584, 48

\bibitem[]{kara}
Karachentsev, I.Makarova, L., ,Makarov, D., Tully, R.,
Rizzi, L., 2015, MNRAS, 337, L85

\bibitem[]{knede}
Khedekar, S., Churazov, E., Kravtsov, A. et al.,
2013, MNRAS, 431, 954

\bibitem[]{kirby 2014}
Kirby, E., Bullock, J., Boylan-Kolchin, M., Kaplinghat, M.,
Cohen, J., 2014, MNRAS, 439, 101

\bibitem[]{kirby 2017}
Kirby, E., et al. 2017, ApJ, 834, 9

\bibitem[]{klypin 1999}
Klypin, A. Kravtsov, A., Valenzuela. O., Prada, F., 1999,
ApJ, 522, 82

\bibitem[{Klypin, A.}]{klypin 2015}
Klypin, A. Karachentsev, I., Makarov, D., Nasonova, O., 2015,
MNRAS, 454, 1798

\bibitem[{Klypin, A.}]{klypin 2011}
Klypin, A. Trujillo-Gomez, S., Primack, J., 2011, ApJ, 740, 102;

\bibitem[]{komatsu-11}
Komatsu, E., et al. 2011, ApJS, 192, 18

\bibitem[]{koposov-09}
Koposov, S., Yoo, J., Rix, H.W., Weinberg, D., Maccio, A.,
Escude, J., 2009, ApJ, 696, 2179

\bibitem[]{kop-13}
Kopp, J., Machado, P., Maltoni, M., Schwetz, T., 2013, JHEP, 05, 050

\bibitem[]{kormendy 2015}
Kormendy, J., \& Freeman, K., 2015, IAUS, 311, 72,

\bibitem[]{kravtsov 2012}
Kravtsov, A., Borgani, S., 2012, ARA\&A, 50, 353

\bibitem[]{kus 2009}
Kusenko, A., Phys.Rep.,2009, 481, 1

\bibitem[]{kuzio 2008}
Kuzio de Naray, R., Mcgaugh, S., de Blok, W., 2008, ApJ,
676, 920

\bibitem[]{lacey}
Lacey, C., \& Cole, S., 1993, MNRAS, 262, 627

\bibitem[]{laport-13}
Laporte, C.,. Walker, M., Penarrubia, J., 2013, MNRAS,
433, 54L

\bibitem[]{larsen-11}
Larson, D.., et al. 2011, ApJS, 192, 16

\bibitem[]{Lib-14}
Libeskind, N., Di Cintio, A., Knebe, A., et al.,
2013, PASA, 30, 39

\bibitem[]{Lithwick-11}
Lithwick Y., Dalal N., 2011, ApJ, 734, 100L

\bibitem[]{Liu}
Liu, C., Peng, E., Cote, P., et al., 2015, ApJ, 812, L34

\bibitem[]{Lloyd-davies}
Lloyd--Davies, E., 2011, MNRAS, 418, 14

\bibitem[]{Ludlow-13}
Ludlow, A. Navarro, J., Boylan-Kolchin, M.,
et al., 2013, MNRAS, 432, 1103L

\bibitem[]{Ludlow-16}
Ludlow, A. Bose, S., Angulo, R., et al. 2016, MNRAS, 460, 1214

\bibitem[]{maccio 2012}
Maccio, A., Paduroiu, S., Anderhalden, D., Schneider, A.,
Moor, B., 2012, MNRAS, 424, 1105

\bibitem[]{macci 2013}
Maccio A., Ruchayskiy O., Boyarsky A., Munos--Cuartas J.,
2013, MNRAS, 428, 882

\bibitem[]{makarov 2011}
Makarov, D., Karachentsev, I., 2011, MNRAS, 412, 2498

\bibitem[]{Mamon-15}
Mamon, G., Chevalier, J., Romanovsky, A., Wojtak, R., 2015,
IAUS, 311, 16,

\bibitem[]{Mangano}
Mangano G., Miele, G., Pastor, S., Pinto, T.,
Pisanti, O., Serpico, P., 2006, NuPhB, 756, 100

\bibitem[]{Mantz}
Mantz, A., Allen, S., Morris, R., Rapetti, D., Applegate, D.,
Kelly, P., von der Linden, A.,Schmidt, R., 2014, MNRAS, 440, 2077

\bibitem[]{marc 2013}
Marcovi$\breve{c}$, K., Viel, M., 2014, PASA, 31, 6

\bibitem[]{marsh 2014}
Marsh, D., Silk, J., 2014, MNRAS, 437, 2652

\bibitem[]{martinez}
Martinez-Delgado, D., Lasker, R., Sharina, M., et al.,
 2016, AJ, 151, 96

\bibitem[]{marsh 2014}
Mayet, F., Green, A., Battat, J., et al., 2016, PhR, 627, 1

\bibitem[]{McCon}
McConnachie, A,, 2012, ApJ, 144, 4

\bibitem[]{McDonald}
McDonald, M., Benson, B. A., Vikhlinin, A., et al.,
2013, ApJ, 774, 23

\bibitem[]{McLeod}
McLeod, D., McLure, R., Dunlop, J., et al.,
2015, MNRAS, 450, 303

\bibitem[]{meik}
Meiksin, A., White, M., Peacock, J., 1999, MNRAS, 304, 851

\bibitem[]{merrit}
Merrit, A., van Dokkum, P., Abraham, R., 2014, ApJ, 787, L37

\bibitem[]{mert}
Merten, J., Meneghetti, M.; Postman, M., et al., 2015,
ApJ, 806, 4

\bibitem[]{dor 2007}
Mikheeva, E., Doroshkevich A., Lukash, V.,, 2007, NCimB,
122, 1393

\bibitem[]{miller}
Miller, S., Ellis, R., Newman A., Benson, A., 2014, ApJ, 782, 115

\bibitem[]{mirri-15}
Mirizzi, A., Mangano, G., Pisanti, O., Saviano, N., 2015,
PhRvD.91.025019

\bibitem[]{mitra}
Mitra, S., Choudhury, T.R., Ferrara, A., 2015, MNRAS, 454, L76

\bibitem[]{moor}
Moore, B., Ghigna, S., Governato, F. et al., 1999, ApJ, 524, L19

\bibitem[]{Moughan-11}
Moughan, B., Giles, P., Randall, S., Jones, C., Formen, W.,
2012, MNRAS, 421, 1583

\bibitem[]{nagai-07}
Nagai, D., Kravtsov, A., Vikhlinin A., 2007, ApJ, 668, 1

\bibitem[]{NFW-95}
Navarro~J.F., Frenk~C.S., \& White~S.D.M., 1995, MNRAS, 275, 720

\bibitem[]{NFW-96}
Navarro~J.F., Frenk~C.S., \& White~S.D.M., 1996, ApJ, 462, 563

\bibitem[]{NFW-97}
Navarro~J.F., Frenk~C.S., \& White~S.D.M., 1997, ApJ, 490, 493

\bibitem[]{Nes-15}
Nesseris, S.,  Sapone, D., 2015, IJMPD, 24, 1550045

\bibitem[]{New-10}
Newman, A., Treu, T., Ellis, R., Sand, D., 2013, ApJ, 765, 25

\bibitem[]{Nipot-15}
Nipoti, C., \& Binney, J., 2015, MNRAS, 446, 182

\bibitem[]{Peebles-67}
Partridge R.B., \& Peebles P.J.E., 1967a, ApJ, 147, 868

\bibitem[]{Peebles-67}
Partridge R.B., \& Peebles P.J.E., 1967b, ApJ, 148, 377

\bibitem[]{Peebles-67}
Peebles P.J.E., 1967, ApJ, 147, 859

\bibitem[]{Peebles-74}
Peebles P.J.E., 1980, The Large-Scale Structure of the
Univers, (Princeton: Princeton Univ. Press)

\bibitem[]{Penar-08}
Penarrubia, J., Navarro, J., McConnachie, A., 2008, ApJ, 673,
226

\bibitem[]{Penar-10}
Penarrubia, J., Benson, A., Walker, M., et al.,
2010, MNRAS, 406, 1290

\bibitem[]{Penar-12}
Penarrubia, J., Pontzen, A., Walker, M., Koposov, S.,
2012, ApJ, 759L, 42

\bibitem[]{Pili}
Pilipenko, S., Doroshkevich, A., Lukash, V., Mikheeva, E.,
2012, MNRAS, 427L, 30

\bibitem[]{pointe_05}
Pointecouteau, E., Arnaud, M., Pratt, G., 2005, A\&A, 435, 1

\bibitem[]{pont}
Pontzen, A., \& Governato, F., 2014, Nature, 506, 171,

\bibitem[]{Prada 2012}
Prada, F., Klypin, A., Cuesta, A., et al.,
2012, MNRAS, 423, 3018

\bibitem[]{pratt_06}
Pratt, G., Arnaud, M., Pointecouteau, E., 2006, A\&A, 446, 429

\bibitem[]{pratt_09}
Pratt, G., Croston, J., Arnaud, M., B\"oringer, H., 2009,
A\&A, 498, 361

\bibitem[]{pratt_10}
Pratt, G., Arnaud, M., Piffaretti, R., et al., 2010, A\&A, 511, 85

\bibitem[]{press_74}
Press, W., Schechter, P., 1974, ApJ, 187, 425

\bibitem[]{rein-2013}
Reichardt, C., Stadler, B., Bleem, L., et al., 2013, ApJ, 763, 127

\bibitem[]{rob-2015}
Robertson, B., Ellis, R., Furlanetto, S., Dunlop, J.,
2015, ApJ, 802, L19

\bibitem[]{rub-2014}
Rubakov, V., 2014, Phys. Usp. 57 128

\bibitem[]{ru-2013}
Ruel J., Bazin, G., Bayliss, M., et al., 2014, ApJ, 792,45

\bibitem[]{ruff-2015}
Ruffini  R., Arguelles, C., Rueda, J., 2015, MNRAS, 451, 622

\bibitem[]{salucci 2015}
Saliwanchik, B., Montroy,T., Aird, K., et al. 2015, ApJ, 799, 137

\bibitem[]{sam 2014}
Samushia, L., Reid, B., White, M., et al., 2014, MNRAS, 439, 3504

\bibitem[]{saro 2014}
Saro, A., Liu, J., Mohr, J., et al., 2014,MNRAS, 440, 2610

\bibitem[]{save 2013}
Savelainen, M., Valiviita, J., Walia, P., et al.,
2013, PhRvD., 88, 063010.

\bibitem[]{sawa 2013}
Sawala, T., Frenk, C., Crain, R., et al.,
2013, MNRAS, 431, 1366

\bibitem[]{sawa 2016}
Sawala, T., Pihajoki, P., Johannsson, P., et al., 2016, arXiv:160905214;
Sawala, T., Frenk, C., Fattahi, A., et al., 2016, MNRAS, 457, 1931
Sawala, T., Frenk, C., Fattahi, A., et al., 2016, MNRAS, 456, 85

\bibitem[]{schev 2015}
Schewtschenko, J., Baugh, C., Wilkinson, R., et al. 2016,
MNRAS, tmp, 862; 

\bibitem[]{schnei 2013}
Schneider, A., Smith, R., Reed, D., 2013, MNRAS, 433, 1573

\bibitem[]{schn 2013}
Schneider, A., Anderhalden, D., Maccio, A., Diemand, J.,
2014, MNRAS, 441, L6

\bibitem[]{schu 2014}
Schultz, C., Onorbe, J., Abazajian, K., Bullock, J., 2014,
MNRAS, 442, 1597

\bibitem[]{sheth}
Sheth, R., Tormen, G., 2002, MNRAS, 329, 61;
2004, MNRAS, 350, 1385

\bibitem[]{suhada}
Suhada, R., et al., 2012, A\&A, 537, 39, 1076,

\bibitem[]{Tasit-02}
Tasitsiomi, A., Kravtsov, A., Gottl\"ober, S., Klypin, A.,
2004, ApJ, 607, 125

\bibitem[]{Tess.}
Teyssier, R., Pontzen, A., Dubois, Y., Read, J., 2013,
MNRAS, 429, 3068

\bibitem[]{Tollerud-12.}
Tollerud, E., Beaton, R., Geha, M.,  et al., 2012, ApJ, 752, 45

\bibitem[]{Tollerud-14.}
Tollerud E., Boylan-Kolchin, M., Bullock, J., 2014, MNEAS, 440, 3511

\bibitem[]{trujillo}
Trujillo-Gomez, S., Klypin, A., Primack, J., Romanowsky, A., 2011,
ApJ, 742, 16

\bibitem[]{turner 1984}
Turner, M., Steigmann, G., Krauss, L., 1984, Phys.Rev.Let.,
52, 2090.

\bibitem[]{umetsu}
Umetsu, k., Medezinski, E., Nonino, M., et al., 2014, ApJ, 795, 163

\bibitem[]{dokkum-15}
van Dokkum, P., Abraham, R., Merrit, A., et al.,
2015, ApJ. 798, L45

\bibitem[]{dokkum}
van Dokkum, P., Abraham, R., Brodie, J., et al.,
2016, ApJ. 828, L6

\bibitem[]{viel 2013}
Viel, M., Becker, G., Bolton, J., Haehnelt,M., 2013,
PhRvD, 88d3502

\bibitem[]{vikhlinin-06}
Vikhlinin, A., Kravtsov, A., Forman, W., et al.
2006, ApJ, 640, 691

\bibitem[]{vikhlinin}
Vikhlinin, A., Burenin, R., Ebeling, H., et al., 2009, ApJ, 692, 1033

\bibitem[]{walker 2009}
Walker, M., Mateo, M., Olszewski, E., et al. 2009, ApJ., 704, 1274

\bibitem[]{walk 2012}
Walker, M., Penarrubia, J., 2011, ApJ,  742, 20

\bibitem[]{wang 2013}
Wang, M.-Y., Croft, R., Peter, A., Zentner, A., Purcell, C.,
2013, PhRvD, 88l3515

\bibitem[]{wang 2014}
Wang, M-Y., Peter, A., Strigari, ., Zentner, A., Arant, B.,
Garrison-Kimmel, S., Rocha, M., 2014, MNRAS, 445, 614

\bibitem[]{weisz 2014a}
Weisz, D., et al., 2014a, ApJ, 789, 24

\bibitem[]{weisz 2014b}
Weisz, D., et al., 2014b, ApJ, 789, 147

\bibitem[]{wetzel 2015}
Wetzel, A., Deason, A., Garrison -- Kimmel, S., 2015, ApJ, 807, 49

\bibitem[]{whitt 2014}
Whittaker, L., Brown, M., Battye, R., 2014, MNRAS, 445, 1836

\bibitem[]{whitt 2015}
Whittaker, L., Brown, M., Battye, R., 2015, MNRAS, 451, 383

\bibitem[]{Zel'd-70}
Zel'dovich Ya.B.,  1970, A\&A, 5, 84

\bibitem[]{Zel'd}
Zel'dovich Ya.B., Novikov I.D., 1983, Structure and evolution
of the Universe, University of Chicago Press.

\bibitem[]{zhang 2006}
Zhang, Y., B\"oringer, H., Finoguenov, A., et al.,
2006, A\&A, 456, 55

\bibitem[]{zitr}
Zitrin, A., Labbe, I., Belli, S., et al., 2015, ApJ, 810, L12

\bibitem[]{zysina}
Zysina, N., Fomichev, S., Khruschov, V., 2014, PAN, 77, 890
\end{thebibliography}

\end{document}